\documentclass[]{agujournal2019}
\usepackage{lmodern}
\usepackage{lineno}
\usepackage[finalnew]{trackchanges}
\usepackage{soul}
\draftfalse

\journalname{JGR: Solid Earth}

\usepackage{xr}
\begin{document}


%
%


\title{Tectonics on early Earth driven by Earth's changing shape during tidal recession of the Moon}

%
%




\authors{S. J. Lock\affil{1,2}}

\affiliation{1}{School of Earth Sciences, University of Bristol, Bristol, BS8 1RJ, UK}
\affiliation{2}{Formerly: Division of Geological and Planetary Sciences, California Institute of Technology, Pasadena, CA 91125, USA}




\correspondingauthor{S. J. Lock}{s.lock@bristol.ac.uk}



\begin{keypoints}
\item Earth's first crust would have been thicker at the equator than at the poles due to the planet's rapid rotation
\item Change in Earth’s shape as its day lengthened provided a strong driver for tectonics during the first millions to tens of millions of years
\item Such tectonics could have led to production of more-felsic rocks and chemically-distinct mantle domains shortly after Moon formation
\end{keypoints}

%
%

%
%


\begin{abstract}
Few fragments of Earth's earliest crust have survived, and so this formative period of our planet's history is shrouded in mystery. To leverage the limited data available, it is important to understand all the processes that could have shaped our young planet. Here, I explore a previously neglected aspect of early-Earth dynamics: rotation. Earth after the Moon-forming impact was rapidly rotating with its rotation rate slowing significantly during the first tens of millions of years after the impact as the Moon tidally receded. Using planetary structure and tidal evolution calculations, I show that the initially distorted structure of Earth would likely have led to significant latitudinal variations in primary crustal thickness. Further, the subsequent changes in Earth's shape in response to its slowing rotation rate could have driven extensive tectonic activity and modulated the response of the crust to other forces. There was extension of the crust in polar regions and convergence in the equatorial regions at rates potentially comparable to those forming the Himalayas today\add{ in high-angular momentum scenarios}. Depending on the rheology of the earliest crust, such motions could have led to burial of hydrated and weathered surface material, extensive volcanism, and degassing of Earth's interior. Phase changes due to increases in internal pressure could have led to delamination of parts of the crust, producing early mantle heterogeneities. Hydrated secondary melting of the crust and the production of evolved magmas could explain the emergence of a diversity of rock types within tens of millions of years of Earth’s formation. 
\end{abstract}

\section*{Plain Language Summary}

Just after the Earth and Moon formed, the Moon was much closer to Earth than it is now, and Earth was rapidly rotating with a day shorter than about five hours. Earth spinning that fast would have been somewhat flattened, in the most extreme cases with the equator being twice as far away from the center of the planet as the poles. Over Earth's history, the gravitational tug between Earth, the Moon, and the Sun has led to the Moon moving away from Earth, Earth's day becoming longer, and Earth becoming less flattened and more spherical. Earth's first crust formed whilst the planet was still rapidly rotating and this would have caused the crust to be thicker in the equator than at the poles, by as much as tenfold. As Earth's rotation rate slowed, the change in its shape would have deformed the crust, squishing it together nearer the equator and pulling it apart at the poles. This turmoil would have shaped the atmosphere and crust of Earth and helped form an environment that was suitable for the emergence and survival of life.

%
%

%


%
%
%
%

\section{Introduction}

The properties and dynamics of the crust of early Earth controlled many key aspects of the Earth system at a formative stage in its evolution, including the availability of different elements in surface environments, topography and the potential for dry land, and the composition of the atmosphere through volcanism and erosion. However, destruction of almost all of the geological record from the first few hundred million years of Earth's history by the subsequent billions of years of tectonics and erosion makes it challenging to infer much about the ancient crust and early surface environment of Earth. In order to make the most of the limited observational constraints available to elucidate the surface conditions and dynamics of early Earth, it is essential to have a robust awareness and understanding of all the geophysical processes that governed the growth and evolution of the early crust \cite{ONeil2019_early_crust_book,Korenaga2023_rapid_solidification,korenaga_hadean_2021,Davies1992,Solomatov2015,Abe1993}.

Earth's first long-lived crust emerged in the aftermath of the Moon-forming giant impact \cite{Hartmann1975,Cameron1976}, which marked the end of the main stage of Earth's formation. Another planet-sized body hit the proto-Earth, injecting material into orbit out of which the Moon was formed, and melting and vaporizing much of the silicate Earth \cite{Nakajima2015,Lock2017,postema_thermal_nodate,postema_thermal_2026}. Radiation from the edge of the post-impact structure, and later the top of Earth's atmosphere, led to condensation and solidification of the mantle and surface within a couple of million years, forming the first terrestrial crust \cite{Elkins-Tanton2008,Lebrun2013,Zahnle2015,Korenaga2023_rapid_solidification,Lock2020}. The petrology, thickness, and stability of this first crust is highly debated \cite{Carlson2019,Harrison2017,Griffin2014}, as is the mode of mantle convection that regulated its subsequent evolution \cite{korenaga_hadean_2021,ONeill2018,Carlson2019,al_asad_coupled_2024}. There are therefore a range of geophysical models that attempt to match different geochemical and petrological constraints on the formation and evolution of Earth's earliest crust \cite<e.g.,>{Moore2013,Gerya2015,Rozel2017,Miyazaki2022_heterogenious_mantle,Marchi2014}.


However, previous work has neglected an important component of early Earth dynamics: rotation. When the Moon first formed, Earth was rotating very rapidly. The huge torques exterted during the Moon-forming giant impact left the Earth-Moon system with considerable angular momentum \cite{Canup2001,Cuk2012,Canup2012,Lock2017,Rufu2017,Lock2018moon,Reufer2012,meier_origin_2025,timpe_systematic_2023}. After the impact, the Moon formed between 2.9 and about 6 Earth radii ($R_{\rm Earth}$) from the center of Earth \cite{Kokubo2000,Salmon2012,Lock2018moon}, much closer than at present (today the lunar semi-major axis, $a_{\rm Moon}\sim\,$60~$R_{\rm Earth}$). Consequently, more of the angular momentum of the Earth-Moon system was accommodated in Earth's rotation than at present, and Earth was rotating rapidly. In addition, the total angular momentum of the Earth-Moon system could have been substantially larger when the Moon first formed, if angular momentum \change{has been}{was} subsequently transferred to the Earth-Sun system during lunar tidal evolution \cite{Cuk2012,Wisdom2015,Cuk2016,Tian2020,Ward2020,Rufu2020,Cuk2021}. Tidal evolution models have shown that the initial angular momentum of the Earth-Moon system could have been more than twice its present-day value ($L_{\rm EM}=3.5 \times 10^{34}$~kg~m$^2$~s$^{-1}$), and recent Moon-formation models have considered scenarios with a range of initial angular momenta as high as about 3~$L_{\rm EM}$ \cite{Cuk2012,Canup2012,Lock2018moon,meier_systematic_2024,timpe_systematic_2023}. I will refer to the scenario where the angular momentum of the Earth-Moon system has been constant since Moon formation as the `canonical model', and those scenarios in which the angular momentum of the system was initially greater than today as `high-angular momentum models'. As a result of its higher angular momentum, when the Moon first formed Earth had a length of day between \change{2.3 and about 5 hrs}{about 5 (in the canonical model) and 2.3 hrs (for high angular-momentum models)}.

\begin{figure}
\centering
\includegraphics{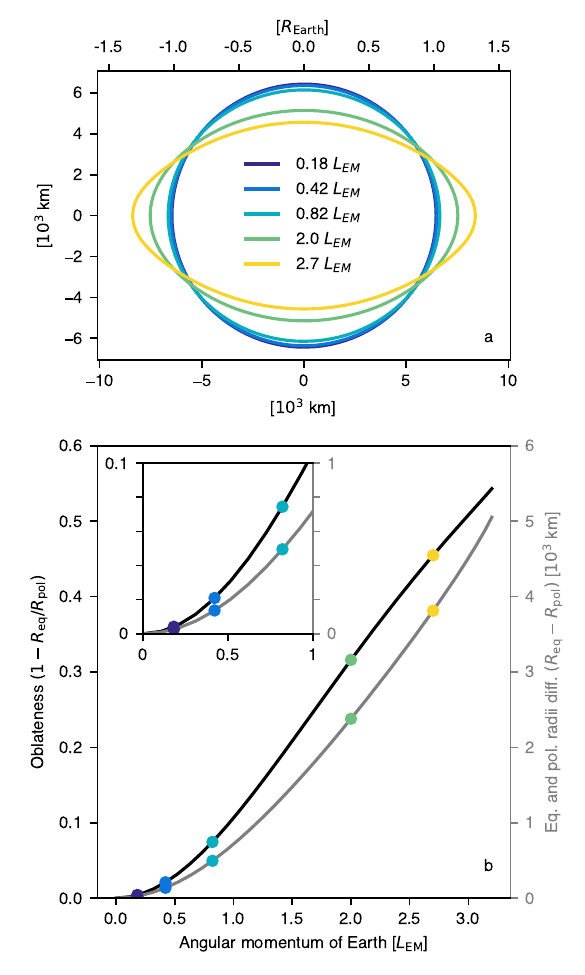}
\caption{Caption on next page.}
\end{figure}
\addtocounter{figure}{-1}
\begin{figure}
\centering
\caption{\add{The shape of Earth with different angular momentum representing different stages of lunar recession, demonstrating that }the early Earth was substantially oblate and its shape changed considerably during lunar tidal evolution. a: Colored lines are the outlines of Earth with different angular momenta: 0.18~$L_{\rm EM}$, the angular momentum of Earth today; 0.42~$L_{\rm EM}$, the angular momentum of Earth at the Cassini state transition; 0.82~$L_{\rm EM}$, the angular momentum of Earth when the Moon was at the Roche limit ($2.9$~$R_{\rm Earth}$) in the canonical model, the maximum post-Moon formation angular momentum of Earth in this scenario; 2~$L_{\rm EM}$, \change{similar to the angular momentum of Earth at the start of the simulations shown in Figures~7 and S2--S4 when the Moon was close to the Roche limit}{roughly the angular momentum of Earth when the Moon was at the Roche limit in the simulations of high angular momentum scenarios shown in Figures}~\ref{fig:CukFig7_PTcomp} and \ref{sup:fig:RufuA5_PTcomp}--\ref{sup:fig:RufuA10k_PTcomp}; and 2.7~$L_{\rm EM}$, typical of the angular momentum of Earth when the Moon was at the Roche limit after the highest angular momentum Moon-forming impacts
\cite{Cuk2012,Canup2012,Lock2018moon}. b: The oblateness (black line, left axis) and difference between equatorial ($R_{\rm eq}$) and polar ($R_{\rm pol}$) radii (grey line, right axis) for Earth as a function of angular momentum. Colored dots correspond to the example bodies shown in panel a. Inset is a close up of lower angular momenta.
}
\label{fig:shape}
\end{figure}

The rapid rotation of early Earth forced the planet to be substantially oblate (Figure~\ref{fig:shape}) and the pressures in Earth to be considerably lower than at the present day \cite{Lock2019pressure,Lock2020}. In the highest angular momentum Moon-formation models \cite{Cuk2012,Canup2012,Lock2018moon,meier_systematic_2024,timpe_systematic_2023}, Earth could have had an equatorial radius twice that of its polar radius (e.g., 2.7~$L_{\rm EM}$ example in Figure~\ref{fig:shape}) and a core-mantle boundary (CMB) pressure less than half that of the present day \cite{Lock2019pressure,postema_thermal_nodate,postema_thermal_2026}. 

Over time, tidal interactions between Earth and the Moon drove the lunar orbit outwards and slowed Earth's rotation, leading to a substantial change in Earth's shape and structure \cite{Lock2019pressure,Lock2020}. The dynamics and timescale of lunar tidal recession are highly debated due to uncertainties in the tidal properties of hot, partially-molten bodies \cite{Henning2014,Korenaga2025_tidal_3_compaction}, the potential for strong feedbacks between tidal dissipation and thermal structure \cite{Zahnle2015}, and the possibility of resonances, near-resonance interactions, and dynamical instabilities \cite{Touma1998,Cuk2012,Wisdom2015,Cuk2016,Ward2020,Tian2020,Rufu2020,Cuk2021}. When the Moon was at about 30~$R_{\rm Earth}$ the lunar orbit went through the Cassini state transition, where the current configuration of the lunar rotation axis, the lunar orbital plane, and the ecliptic plane was obtained \cite{Ward1975}. At the Cassini state transition Earth had an angular momentum of around 0.4~$L_{\rm EM}$ and was only moderately oblate, with a shape and pressure structure relatively similar to the present day (Figure~\ref{fig:shape}). There is no known mechanism that is expected to have substantially changed the angular momentum of the Earth-Moon system after the Cassini state transition and so the transition provides a useful reference point in orbital evolution. 

There are considerable uncertainties on the relative timing of magma ocean solidification, and hence primary crust formation, after the Moon-forming giant impact and the orbital evolution of the Moon. With very few direct observational constraints on the lunar orbit, estimates of the time at which the Cassini state transition occurred vary from 10$^6$--10$^8$~yrs \cite{Dwyer2011,Zahnle2015,Chen2016,Qin2018,Cuk2012,Wisdom2015,Cuk2016,Cuk2021}, depending largely on the highly uncertain tidal properties of the early Earth and Moon. \add{Recently, it has been argued, on the basis of the need for a substantial tidal heating event to explain the ages of lunar samples, that the Moon did not undergo the Laplace plane transition (which occurs before the Cassini-state transition when the Moon is about 20~$R_{\rm Earth}$ from Earth) until $\sim150$~Myr after the formation of the Moon }\cite{nimmo_tidally_2024,tian_investigating_2026}\add{.} The timescale for solidification of \change{the mantle}{Earth's surface} after the Moon-forming impact is similarly uncertain, dictated by assumptions of atmospheric properties and crystal dynamics. However, in \change{these models}{the majority of models of magma ocean solidification} the surface of Earth solidifies, and hence the crust is formed, within at most a couple of million years\add{ of Moon formation}, although some melt can persist deeper in the mantle for much longer \cite{Elkins-Tanton2008,Lebrun2013,Zahnle2015,Korenaga2023_rapid_solidification,Salvador2017,Hamano2013}. It is therefore probable that the first crust formed when the Moon was still close to Earth \cite{Korenaga2023_rapid_solidification} and substantially oblate, and that the crust would have been deformed as Earth's shape changed during tidal evolution\add{ (I discuss this topic further in Section}~\ref{sec:discussion:formation}\add{)}. 


Here, I use planetary structure and tidal evolution models to determine the conditions of early crust formation and how the crust could have been deformed during lunar tidal recession. Although previous work has considered the effect of the change of shape in response to changes in rotation rate on smaller solar system bodies \cite{JayMelosh1977,Dobrovolskis1982,Matsuyama2008,Matsuyama2010,Rhoden2012}, the possibility that such a process could have occurred on early Earth has not been explored. Furthermore, existing work depends on various assumptions, such as small deformation from a sphere and/or constant density bodies, that do not apply to the early Earth. The purpose of this work is to quantify the effect of Earth's evolving rotation following different Moon-formation scenarios and explore its implications for the formation and evolution of Earth's crust and the early terrestrial surface environment. Future work will develop the techniques necessary to quantify the geophysical and geochemical processes on early Earth in more detail. 

This paper begins with an outline of the methods (Section~\ref{sec:methods}), description of the initial state of the proto-Earth as the first crust formed (Section~\ref{sec:results:environment}), and quantification of the surface deformation during lunar tidal recession for both canonical and high angular-momentum lunar formation models (Section~\ref{sec:results:recession}). In Section~\ref{sec:discussion}\add{,} I then discuss the implications of the results for early Earth and conclude in Section~\ref{sec:conclusion}. \ref{sup:sec:tidal_model:dependence} contains a detailed scaling analysis for deformation of the crust during lunar tidal recession in the canonical scenario, \ref{sup:sec:tidal_params}\add{ gives the equivalence of parameters used in different tidal models, }and supplementary information accompanies the manuscript that includes additional examples of high-angular momentum Moon-formation scenarios, animations of several figures, and the scripts and data required to reproduce the work.

\section{Methods}
\label{sec:methods}

I combined planetary structure calculations to model the equilibrium structure of Earth at different times in lunar tidal evolution (Section~\ref{sup:sec:HERCULES}) with tidal evolution models to calculate the strain of the surface of Earth in the canonical (Section~\ref{sup:sec:tidal_model}) and various high-angular momentum (Section~\ref{sup:sec:highAM}) Moon-formation scenarios. These models\remove{, and the connections between them,} are described below.

\subsection{Planetary structure calculations}
\label{sup:sec:HERCULES}

Earth's structure at different stages in lunar tidal evolution was calculated using the open-source HERCULES planetary structure code version 1.0 \cite{Lock2017, Lock2019HERCULES}. In HERCULES, a body is modeled as consisting of a series of nested concentric layers of constant density with each layer described by a series of points at different angles from the rotation axis, $\theta$. HERCULES then uses a potential field method to calculate the equilibrium structure of the body with a given thermal state, composition, mass and angular momentum, using realistic equations of state. In this work, I modeled an Earth-mass planet with the present-day Earth's core mass fraction of 0.323 \cite{Yoder1995}. The planetary structure was calculated at angular momentum steps of 0.1~$L_{\rm EM}$ with the properties at intermediate angular momenta determined by linear interpolation. The mantle and core were assumed to be forsterite and pure iron, respectively, and were modeled using the the ANEOS equation of state (EOS) model \cite{thompson1972,Canup2012,Melosh2007}. The EOS are documented as `aneos-T70' (iron) and `aneos-gadget' (forsterite) respectively in two Zenodo repositories \cite{Stewart2020ironEOS,Stewart2019forsteriteEOS}. The mantle was assumed to be isentropic with a specific entropy of 3.2~kJ~K$^{-1}$~kg$^{-1}$ corresponding to a mantle potential temperature of around 1900~K. This thermal state approximates that of the Hadean mantle \cite{Sleep2010,korenaga_hadean_2021,ONeill2014}. The core was also assumed to be isentropic and have a thermal state similar to the present day, following the arguments of \citeA{Andrault2016}\add{ that the thermal state of Earth's core has changed little since solidification of the lower mantle}. The core specific entropy was set at 1.5~kJ~K$^{-1}$~kg$^{-1}$, corresponding to a temperature of 3800~K at the pressure of the present-day core-mantle boundary. \add{Note that the choice of EOS and (any reasonable) thermal state have only minor effects on our results as the percent level differences in material density between solid and liquid EOSs for major Earth materials }\cite<e.g.,>{Stewart2019forsteriteEOS,Stixrude2005thermo} \add{are small compared to the first-order mass distribution changes in Earth as a consequence of changes in rotation rate.} 

I used the same HERCULES parameters as in previous studies, that have been shown to accurately model the structure of Earth-like planets \cite{Lock2017,Lock2019pressure,Lock2019HERCULES,Lock2020}. The planets were modeled as consisting of $N_{\rm lay}^{\rm core}=20$ evenly spaced layers in the core and $N_{\rm lay}^{\rm mantle}=80$ layers in the mantle, with $N_{\mu}=1000$ points describing the shape of each layer. The expression for the gravity field was truncated at order $2k_{\rm max}=12$. The minimum pressure at the surface of the planet was set to 10~bar. The tolerance for the convergence of the shape of equipotential layers was $\xi_{\rm toll}^{\mu}=10^{-10}$ and the tolerance for the convergence of the mass of the planet was $\xi_{\rm toll}=10^{-8}$. The step used for calculating gradients in the solution algorithm was $\delta\xi=10^{-2}$. For further details of the definitions of these parameters the reader is referred to the HERCULES user manual \cite{Lock2019HERCULES}. 

The effective gravity at the surface of each planet was determined by calculating the gradient of the combined potential (the sum of gravitational and rotational potential) perpendicular to the surface of the planet. As HERCULES natively calculates the potential on the top of each layer at given latitudes ($\theta$), it is straightforward to calculate the radial component of gravity at the planet's surface, $g_r$, by taking the gradient of the potential field using a second order backwards difference method away from the surface along a line of constant $\theta$. The absolute magnitude of the gravity (i.e., that perpendicular to the surface, $g$) is then given by 
\begin{equation}
    g=\frac{g_{r}}{\cos\left(\theta - \lambda \right )} \, ,
\end{equation}
where 
\begin{equation}
    \lambda=\tan^{-1}\left ( \frac{\mathrm{d}z}{\mathrm{d}r_{xy}} \right ) 
\end{equation}
is the angle of the surface relative to the equatorial plane, $z$ is the height of the surface above the equatorial plane, and $r_{xy}$ is the radius of the surface perpendicular to the rotation axis. $\mathrm{d}z/\mathrm{d}r_{xy}$ was found by taking the gradient of the position of the surface points to second order.

In calculating the deformation of the surface during lunar tidal recession I assumed that the strength of the crust was negligible and so the surface followed a gravitational equipotential. \add{This assumption is motivated by the fact that the stresses induced by even small ($\sim0.01$\%) changes in local length are greater than the yield/shear strength of common crustal rocks, and further much of the planet, even at relatively modest depth, would be hot and so responding viscously on short timescales.} Further, I assumed that the change in shape was accommodated locally, i.e., that the deformation can be determined by calculating the change in the local lengths and area of the surface at a given latitude. The local lengths and area on the surface are 
\begin{equation}
    \delta l_{\rm lat} = \sqrt{r^2+\left (\frac{\mathrm{d}r}{\mathrm{d}\theta} \right )^2} \delta \theta \, ,
\end{equation}
\vspace{2pt}
\begin{equation}
    \delta l_{\rm lon} = r \sin{\theta} \delta \phi \, ,
\end{equation}
and
\begin{equation}
    \delta A = \delta l_{\rm lat} \delta l_{\rm lon} \, ,
\end{equation}
where $\delta l_{\rm lat}$ and $\delta l_{\rm lon}$ are the local lengths in the latitudinal and longitudinal directions respectively, $\delta A$ is the local area, $r$ is the radius of the surface which is a function of latitude, $\phi$ is the longitude, and $\delta \theta$ and $\delta \phi$ are increments of latitude and longitude respectively. The gradient $\mathrm{d}r/\mathrm{d}\theta$ was calculated by second order differentiation of the radius of the surface points. The rate of change of local length (or area) with angular momentum of the Earth, $\mathrm{d}\delta l/\mathrm{d}L_{\rm Earth}$ where $L_{\rm Earth}$ is the angular momentum of Earth, was determined by second order differentiation of the local lengths for planetary structures with different angular momenta. The rate of change of local length (or area) as a function of time, $\mathrm{d}\delta l/\mathrm{d}t$ where $t$ is time, was then calculated by multiplying $\mathrm{d}\delta l/\mathrm{d}L_{\rm Earth}$ by $\mathrm{d}L_{\rm Earth}/\mathrm{d}t$ taken from a specific tidal evolution model.

\subsection{Simple tidal evolution model for the canonical scenario}
\label{sup:sec:tidal_model}
\label{sup:sec:tidal_model:description}

\remove{To determine the crustal deformation rates }In the canonical scenario, I used a simplified lunar tidal evolution model \add{to allow examination of the crustal deformation rates during lunar tidal recession without having to consider the degeneracies of lunar orbital history in, for example, lunar orbital inclination, orbital eccentricity, and the obliquity of Earth and the Moon.} \change{To build the simplest tidal evolution model possible}{To this end,} I neglect tidal interaction with the Sun or other external bodies and assume that the inclination and eccentricity of the lunar orbit, and Earth's obliquity are all negligible. In this limit, the rate of change of the semi-major axis of the lunar orbit, $a$, can be described by
\begin{equation}
\label{sup:eqn:adot}
 \frac{\mathrm{d}a}{\mathrm{d}t}=3\frac{k_{2,{\rm Earth}}}{Q_{\rm Earth}} \frac{M_{\rm Moon}}{M_{\rm Earth}} \left ( \frac{R_{\rm Earth}}{a} \right )^5 a \, n \, ,
 \end{equation}
 where: $t$ is time; $k_{2,{\rm Earth}}$ is the tidal Love number of Earth at time $t$; $Q_{\rm Earth}$ is similarly the instantaneous tidal quality factor of Earth; $M_{\rm Moon}$ and $M_{\rm Earth}$ are the mass of the Moon and Earth, respectively; $R_{\rm Earth}$ is the radius of Earth; $n$ is the mean motion of the lunar orbit, given by
 \begin{equation}
 \label{sup:eqn:n}
     n=\sqrt{\frac{G(M_{\rm Moon}+M_{\rm Earth})}{a^3}} \, ;
 \end{equation}
 and $G$ is the gravitational constant \cite{Murray1999}. 
 
 \add{The key element to determining crustal deformation rates is the rate at which the angular momentum of Earth changed and hence the rate the lunar orbit expanded ($\mathrm{d}a/\mathrm{d}t$), and this simple tidal evolution model captures the first order controls on the rate of tidal recession of the Moon. Even in more complete tidal evolution models that, for example, include the effect of eccentricity and inclination, the linear dependence of $\mathrm{d}a/\mathrm{d}t$ on $k_2/Q$ still holds, albeit the tidal recession rate is dependent on the $k_2/Q$ of both the Earth and Moon, and, the $a^{-11/2}$ dependence of $\mathrm{d}a/\mathrm{d}t$ is still dominant. Only in scenarios with more complex dynamical interactions --- such as resonances }\cite{Cuk2012}\add{, near resonance interactions }\cite{Wisdom2015,Tian2020,Ward2020,Rufu2020}\add{, and instabilities }\cite{Cuk2016,Tian2020,Cuk2021}\add{ --- do other tidal effects become significant (see Section}~\ref{sup:sec:highAM}\add{). In the canonical scenario the system is not expected to undergo any significant resonant interactions or unstable transitions }\cite{Touma1994,Touma1998}\add{ before the Cassini-state transition} \cite{Cuk2016}\add{, and so neglecting any more complex dynamics is a reasonable simplification.}
 
 Equation~\ref{sup:eqn:adot} can be integrated to give an expression for the lunar semi-major axis over time:
 \begin{equation}
 \label{sup:eqn:a}
     a= \left [ \frac{39}{2} \frac{M_{\rm Moon}\sqrt{G(M_{\rm Earth}+M_{\rm Moon})}}{M_{\rm Earth}} R_{\rm Earth}^5 \, \mathcal{I}(t) -a_0^{13/2} \right ]^{2/13} \, ,
 \end{equation}
where $a_0$ is the initial semi-major axis at $t=0$, $\mathcal{I}$ is the integral
\begin{equation}
    \mathcal{I}(t)=\int_{0}^{t} \frac{k_{2,{\rm Earth}}(t')}{Q_{\rm Earth}(t')} dt' \, ,
\end{equation}
and $t'$ is a `dummy' variable of integration. \add{Over the angular momentum range relevant for the canonical model (0.18--0.82~$L_{\rm Earth}$), the change in Earth's radius is small ($\sim$3\%, Figure}~\ref{fig:shape}\add{). I therefore, for simplicity, take $R_{\rm Earth}$ to be constant and equal to that of the present-day Earth. } Earth's angular momentum is given by
 \begin{equation}
 \label{sup:eqn:L}
     L_{\rm Earth} = L_{\rm EM} - M_{\rm Moon} \sqrt{G(M_{\rm Earth}+M_{\rm Moon})a}  \, .
 \end{equation}
 Critical to understanding the rate of deformation of Earth's crust is the rate of change of angular momentum as a function of time. Differentiating Equation~\ref{sup:eqn:L} gives
 \begin{equation}
 \label{sup:eqn:dLdt_adadt}
     \frac{\mathrm{d}L_{\rm Earth}}{\mathrm{d}t} = - \frac{M_{\rm Moon}}{2} \sqrt{\frac{G(M_{\rm Earth}+M_{\rm Moon})}{a}} \frac{\mathrm{d}a}{\mathrm{d}t}  \, .
 \end{equation}
 Using Equations~\ref{sup:eqn:adot} and \ref{sup:eqn:n} this can be expressed as a function of $a$,
 \begin{equation}
 \label{sup:eqn:dLdt_a}
     \frac{\mathrm{d}L_{\rm Earth}}{\mathrm{d}t} = - \frac{3}{2} \frac{k_{2,{\rm Earth}}}{Q_{\rm Earth}} \frac{G(M_{\rm Earth}+M_{\rm Moon})M_{\rm Moon}^2}{M_{\rm Earth}}  R_{\rm Earth}^5 \left(\frac{1}{a}\right)^{6}   \, ,
 \end{equation}
 which can in turn be expressed solely as a function of time using Equation~\ref{sup:eqn:a}.

 \remove{Using a simple tidal model allows examination of the crustal deformation rates during lunar tidal recession without having to consider the degeneracies of lunar orbital history in, for example, lunar orbital inclination, orbital eccentricity, and the obliquity of Earth and the Moon. }

 In this simple model, the tidal evolution of the Moon is dependent only on the mass and radius of the Earth and Moon, the initial semi-major axis, and the ratio $k_{2,{\rm Earth}}/Q_{\rm Earth}$ which describes the tidal response of Earth. The tidal recession rate is a strong function of semi-major axis ($a^{-11/2}$) and the higher $k_{2,{\rm Earth}}/Q_{\rm Earth}$ the faster the lunar orbit evolves. For reference, the present-day Earth has $k_{2,{\rm Earth}}\sim$0.3, $Q_{\rm Earth}\sim$12, and thus $k_{2,{\rm Earth}}/Q_{\rm Earth}\sim$0.025 \cite{Yoder1995,Egbert2000}. The largest contribution to \change{$Q_{\rm Earth}$}{tidal dissipation} at the present day comes from the oceans and it has been estimated that the tidal quality factor of solid body tides alone is much higher, $Q_{\rm Earth} \sim$100 \cite{Ray2012}. Immediately after the impact, Earth was a mostly fluid structure \cite{Nakajima2015,Lock2017,postema_thermal_nodate,postema_thermal_2026} and likely had a tidal quality factor more equivalent to the gas giants ($Q_{\rm Earth}\sim$10$^{3}$--10$^{4}$) than to a solid planet \cite{Zahnle2015}. Over time, as Earth condensed and solidified, the planet's ability to dissipate tidal energy would have increased and $Q_{\rm Earth}$ would have decreased substantially to be more similar to that of the present-day solid Earth. However, the exact time evolution of Earth's \add{(and the Moon's)} tidal parameters, and the feedback between tidal dissipation and the tidal parameters, is \add{difficult to calculate and }highly uncertain \cite{Henning2014,Zahnle2015,Korenaga2025_tidal_1_inertia_e,Korenaga2025_tidal_2_atmo,Korenaga2025_tidal_3_compaction}. \add{Furthermore, the radius at which the Moon forms can vary from the Roche limit ($\sim2.9$~$R_{\rm Earth}$) to several $R_{\rm Earth}$, introducing additional uncertainty in the orbital history} \cite{Kokubo2000,Salmon2012}\add{.}

\begin{figure}
\centering
\includegraphics{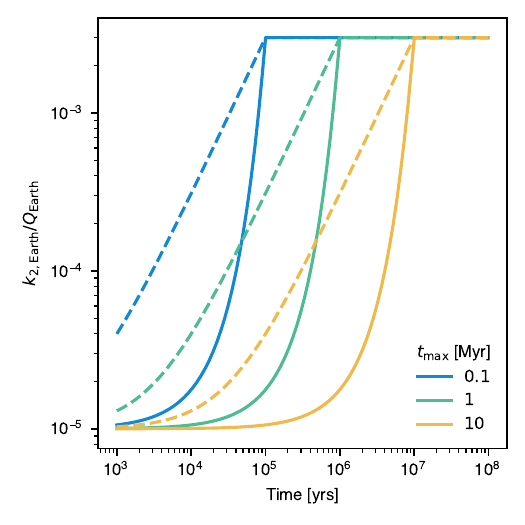}
\caption{Time evolution of $k_{2,{\rm Earth}}/Q_{\rm Earth}$ for exponential (solid lines) and linearly (dashed lines) increasing $k_{2,{\rm Earth}}/Q_{\rm Earth}$ models. Colors indicate the time at which the maximum $k_{2,{\rm Earth}}/Q_{\rm Earth}$ was reached in each scenario ($t_{\rm max}$).
}
\label{sup:fig:k2Q_functions}
\end{figure}
 
 To be able to explore the effect of changing tidal parameters on surface strain, I consider three example time evolution models for $k_{2,{\rm Earth}}/Q_{\rm Earth}$: constant, exponentially increasing, and linearly increasing (Figure~\ref{sup:fig:k2Q_functions}). \add{The scenarios in which $k_{2,{\rm Earth}}/Q_{\rm Earth}$ increases with time mimic the effect of the Earth becoming more dissipative as the mantle freezes (as discussed above) with two different hypothetical evolution pathways.} I stress that \add{any of }these prescribed time evolutions are not intended to be an accurate representation of any particular scenario\add{ or set to reach a particular end state (e.g., the present-day state)}, but rather to allow the effect of, potentially rapidly, increasing tidal dissipation to be explored. When $k_{2,{\rm Earth}}/Q_{\rm Earth}$ is constant,
\begin{equation}
    \mathcal{I}=\frac{k_{2,{\rm Earth}}}{Q_{\rm Earth}} t \, ,
\end{equation}
and
 \begin{equation}
 \label{sup:eqn:a_constk2Q}
     a= \left [ \frac{39}{2} \frac{k_{2,{\rm Earth}}}{Q_{\rm Earth}} \frac{M_{\rm Moon}\sqrt{G(M_{\rm Earth}+M_{\rm Moon})}}{M_{\rm Earth}} R_{\rm Earth}^5 \, t -a_0^{13/2} \right ]^{2/13} \, .
 \end{equation}
 In the second model, $k_{2,{\rm Earth}}/Q_{\rm Earth}$ increases exponentially such that
 \begin{equation}
     \frac{k_{2,{\rm Earth}}(t)}{Q_{\rm Earth}(t)} = \left\{
                \begin{array}{ll}
                  \left. \frac{k_{2,{\rm Earth}}}{Q_{\rm Earth}}\right |_{\rm min} \exp{(\sigma t)} & {\rm for } \; t<t_{\rm max} \\[10pt]
                 \left. \frac{k_{2,{\rm Earth}}}{Q_{\rm Earth}} \right |_{\rm max} & {\rm for } \; t\geq t_{\rm max}
                \end{array}
              \right. 
              \, ,
 \end{equation}
 where
 \begin{equation}
     \sigma = \frac{\ln{\left(\left. \frac{k_{2,{\rm Earth}}}{Q_{\rm Earth}}\right |_{\rm max}\right )}-\ln{\left (\left. \frac{k_{2,{\rm Earth}}}{Q_{\rm Earth}}\right |_{\rm min}\right )}}{t_{\rm max}}
 \end{equation}
 is the exponent defined to ensure continuity between the two time regimes. The integral $\mathcal{I}$ in this case is
 \begin{equation}
     \mathcal{I}(t) = \left\{
                \begin{array}{ll}
                  \left. \frac{k_{2,{\rm Earth}}}{Q_{\rm Earth}}\right |_{\rm min} \left [ \frac{\exp{(\sigma t)}-1}{\sigma} \right ] & {\rm for } \; t<t_{\rm max} \\[10pt]
                 \left. \frac{k_{2,{\rm Earth}}}{Q_{\rm Earth}}\right |_{\rm min} \left [ \frac{\exp{(\sigma t_{\rm max})}-1}{\sigma}  \right ] + \left. \frac{k_{2,{\rm Earth}}}{Q_{\rm Earth}}\right |_{\rm max}  \left (t - t_{\rm max} \right ) & {\rm for } \; t\geq t_{\rm max}
                \end{array}
              \right. \, .
 \end{equation}
 In the third model, $k_{2,{\rm Earth}}/Q_{\rm Earth}$ increases linearly such that
 \begin{equation}
     \frac{k_{2,{\rm Earth}}(t)}{Q_{\rm Earth}(t)} = \left\{
                \begin{array}{ll}
                  \left. \frac{k_{2,{\rm Earth}}}{Q_{\rm Earth}}\right |_{\rm min} + \frac{t}{t_{\rm max}} \left ( \left. \frac{k_{2,{\rm Earth}}}{Q_{\rm Earth}} \right |_{\rm max} -\left. \frac{k_{2,{\rm Earth}}}{Q_{\rm Earth}} \right |_{\rm min} \right ) 
                  & {\rm for } \; t<t_{\rm max} \\[10pt]
                 \left. \frac{k_{2,{\rm Earth}}}{Q_{\rm Earth}} \right |_{\rm max} 
                 & {\rm for } \; t\geq t_{\rm max}
                \end{array}
              \right. \, ,
 \end{equation}
 giving
 \begin{equation}
     \mathcal{I}(t) = \left\{
                \begin{array}{ll}
                  \left. \frac{k_{2,{\rm Earth}}}{Q_{\rm Earth}}\right |_{\rm min} t +\frac{t^2}{2t_{\rm max}} \left ( \left. \frac{k_{2,{\rm Earth}}}{Q_{\rm Earth}}\right |_{\rm max} - \left. \frac{k_{2,{\rm Earth}}}{Q_{\rm Earth}}\right |_{\rm min}\right )
                  & {\rm for } \; t<t_{\rm max} \\[10pt]
                 \frac{t_{\rm max}}{2} \left ( \left. \frac{k_{2,{\rm Earth}}}{Q_{\rm Earth}}\right |_{\rm min} - \left. \frac{k_{2,{\rm Earth}}}{Q_{\rm Earth}}\right |_{\rm max}\right ) + \left. \frac{k_{2,{\rm Earth}}}{Q_{\rm Earth}}\right |_{\rm max} t
                 & {\rm for } \; t\geq t_{\rm max}
                \end{array}
              \right. \, .
 \end{equation}
 \add{In the later two cases, $t_{\rm max}$ is the time at which the transition to a more dissipative regime is completed. In the case that the increased tidal dissipation arises from the solidification of the magma ocean $t_{\rm max}$ would be comparable to the time for freezing the magma ocean (10$^4$-10$^7$~yrs). We have explored a range of $t_{\rm max}$ and minimum and maximum $k_{2,{\rm Earth}}/Q_{\rm Earth}$ values.}


\subsection{High-angular momentum tidal evolution models}
\label{sup:sec:highAM}

In this study I considered two possible mechanisms that have been proposed to transfer angular momentum away from the Earth-Moon system during lunar tidal evolution: instability during the Laplace plane transition due to Earth having an initially high obliquity \cite{Cuk2016,Cuk2021,Tian2020}; and a limit cycle associated with the evection resonance \cite{Wisdom2015}. To calculate the rate of surface deformation in the first of these scenarios I used the results of two simulations shown in Figures~3 and 7 of \citeA{Cuk2021}. In these simulations the tidal evolution is calculated using a `constant-$Q$' approach, where the tidal evolution is controlled by the ratio of the tidal Love number to the tidal quality factor for both the Earth and Moon \cite{Murray1999}. The total angular momentum, initial obliquity, and tidal parameters of the Earth and Moon are different for each example case.

 To calculate the rate of surface deformation in the scenario in which the Earth-Moon system enters a limit cycle, I used the results of the simulations of \citeA{Rufu2020}. \citeA{Rufu2020} performed a suite of lunar tidal evolution simulations with varying tidal parameters but with the same initial Earth-Moon system angular momentum of 2.2~$L_{\rm EM}$. The tidal parameters in the simulation were set by specifying the tidal $A$ parameter\add{ and the tidal time constant, $t_T$ (a translation between these parameters and those used elsewhere is give in }\ref{sup:sec:tidal_params}\add{)}. \remove{, defined as}
\remove{ EQUATION }
\remove{and the tidal time constant, defined as}
\remove{ EQUATION }
\remove{$k_{2, {\rm Earth}}$ and $k_{2, {\rm Moon}}$ are the tidal love numbers of the Earth and Moon, respectively; $\Delta t_{\rm Earth}$ and $\Delta t_{\rm Moon}$ are the terrestrial and lunar tidal time delays, respectively; $M_{\rm Earth}$ and $M_{\rm Moon}$ are the mass of the Earth and Moon; $R_{\rm Earth}$ and $R_{\rm Moon}$ are the radius of the Earth and Moon; $\mu=M_{\rm Moon}/M_{\rm Earth}$; and}
\remove{ EQUATION }
\remove{where $G$ is the gravitational constant. To allow comparison to the other models used in this work, $t_T$ can be related to an effective $k_{2,{\rm Earth}}/Q_{\rm Earth}$ as}
\remove{ EQUATION }
\remove{where $a_0=3.5 R_{\rm Earth}$ and $s_{0, {\rm Earth}}$ are the initial semi-major axis and spin rate of Earth in the simulation.} \citeA{Rufu2020} determined a simulation to be `successful' if the angular momentum of the Earth-Moon system after exiting the limit cycle was between 0.99 and 1.07~$L_{\rm EM}$ \cite<see Figure~6 of>{Rufu2020}. All the examples I consider here meet that condition.

\section{Results}
\label{sec:results}

I will first describe the near-surface pressures and effective gravity that would have been influential during crust formation on a rapidly-rotating Earth (Section~\ref{sec:results:environment}), and then present the results as to how the surface would be strained during tidal recession (Section~\ref{sec:results:recession}).

\begin{figure}
\centering
\includegraphics{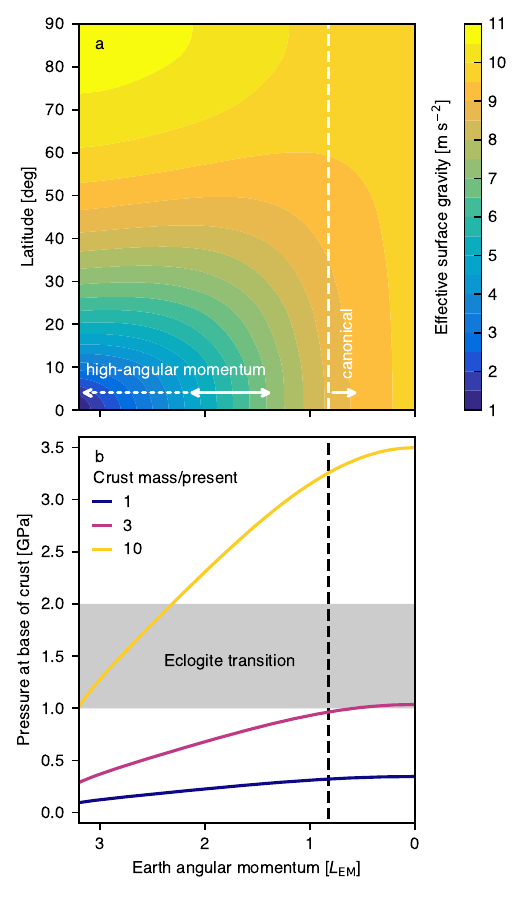}
\caption{Caption on next page.}
\end{figure}
\addtocounter{figure}{-1}
\begin{figure}
\centering
\caption{The effective surface gravity and crustal pressures \add{for Earth with different angular momenta, demonstrating that they} were \add{both typically} much lower on the early Earth. a: The effective gravity perpendicular to the surface at different latitudes on Earth with varying angular momenta. The variation in surface gravity is caused by the variable mass distribution of the oblate planets and the strength of the centrifugal force.  The white dashed line indicates the angular momentum of Earth if the Moon formed at the Roche limit ($\sim$~$2.9 R_{\rm Earth}$) in the canonical scenario, the maximum initial angular momentum of Earth in the canonical scenario. b: The pressure at the base of the crust as a function of angular momentum for crusts of different masses (colored solid lines). Crustal mass is shown relative to the mass of the present-day crust neglecting any depleted mantle produced by crustal production \cite<$2.77\times10^{22}$~kg,>{Mooney1998_crust5_1}. Grey shaded region shows the range of possible pressures for the eclogite transition in modern mid-ocean ridge basalt \cite<MORB, $\sim$1--2~GPa:>{Faccenda2017,Hacker2003_subfac1}. The black dashed line indicates the angular momentum of Earth if the Moon formed at the Roche limit in the canonical scenario. Double-headed white arrows indicate the range of angular momenta explored in tidal (solid) and impact (dashed) studies of high-angular momentum Moon-formation models.}
\label{sup:fig:gravity}
\end{figure}

\subsection{\add{Pressure and
gravity near the surface of a rapidly rotating Earth} and the environment for crustal formation}
\label{sec:results:environment}

Many processes of crust formation, separation, and processing depend on the effective surface gravity on Earth, the net effect of gravitational attraction from the planet\remove{,} and the centrifugal force (see discussion in Section~\ref{sec:discussion:formation}). Figure~\ref{sup:fig:gravity}a shows the magnitude of the effective gravity field perpendicular to the surface (colors) at different latitudes across \change{bodies of}{Earth with} different angular momenta. The gravity on \change{rotating bodies is}{a rapidly rotating Earth would be} much lower at the equator, where the centrifugal force acts to counteract the gravitational attraction of the planet.  As the angular momentum of a body approaches the corotation limit \cite<the maximum angular momentum for a corotating, deforming body;>{Lock2017} the effective gravity at the equator approaches zero. In proposed high-angular momentum Moon-formation scenarios the initial angular momentum of Earth could have been as high as $\sim2.7$~$L_{\rm EM}$ and Earth could have had an equatorial surface gravity as low as $~\sim4$~m~s$^{-2}$ when the Moon was first formed. Even in the canonical case Earth could have had an equatorial gravity of only $\sim8$~m~s$^{-2}$ (white dashed line in Figure~\ref{sup:fig:gravity}a). At the poles of rotating bodies, the effective gravity is somewhat increased over that on a non-rotating planet as the deformed shape means that the polar axis is shorter (Figure~\ref{fig:shape}) and the pole is closer to the concentration of mass in the core and equatorial mantle. Nevertheless, on average the surface gravity would have been lower across early Earth than at the present day.

As a result of the lower surface gravity, the pressure near the surface of the planet \cite<as well as at depth:>{Lock2019pressure,postema_thermal_nodate,postema_thermal_2026} would have been lower than at the present-day. Figure~\ref{sup:fig:gravity}b shows the pressure in Earth at depths equivalent to the base of a crust of different masses\add{, spanning the range of possible crustal masses when including the effect of both rapid rotation and the high mantle potential temperature of early Earth (see Section}~\ref{sec:discussion:formation}\add{),} as a function of angular momenta. For reference, the range of possible pressures for the amphibolite to eclogite transition \cite<$\sim1$--2~GPa:>{Faccenda2017,Hacker2003_subfac1} for mid-ocean ridge basalt is shown as a grey band. The pressure at which the first crust was extracted from the mantle, and then solidified and evolved, would have been a factor of two to three lower than the present-day in high-angular momentum scenarios and several percent lower in the canonical scenario (black dashed line in Figure~\ref{sup:fig:gravity}b).

\subsection{Deformation \add{and evolution }of the crust during lunar tidal recession}
\label{sec:results:recession}

Figure~\ref{fig:total_change} shows the total cumulative change in local lengths and area during lunar tidal recession to the present day as a function of latitude (y-axis) and the initial angular momentum of Earth at the time of Moon formation (x-axis). In the equatorial regions, the local lengths, and hence surface area, decreased as a result of lunar tidal recession. Such compression would have driven, or at least promoted, convergent tectonics. Conversely, poleward of about 20--40$^{\circ}$ (black lines in Figure~\ref{fig:total_change}), north or south, the local surface area increased, resulting in crustal extension. In the canonical scenario, where the angular momentum of the Earth-Moon system has remained constant since its formation, the total change in local surface area is on the order of several percent (panels d--f in Figure~\ref{fig:total_change}), but in high-angular momentum scenarios the cumulative change in local area can be over 100\%. Unlike at tectonic boundaries on Earth today, where strain is roughly uni-directional, strain would have been \add{comparable }in all directions and the principal axis \change{varies}{would have varied} with latitude and the angular momentum of the planet. \add{How such strain would actually be accommodated is uncertain (see }Section~\ref{sec:discussion:response}\add{).}

\begin{figure}
\centering
\includegraphics[width=\textwidth]{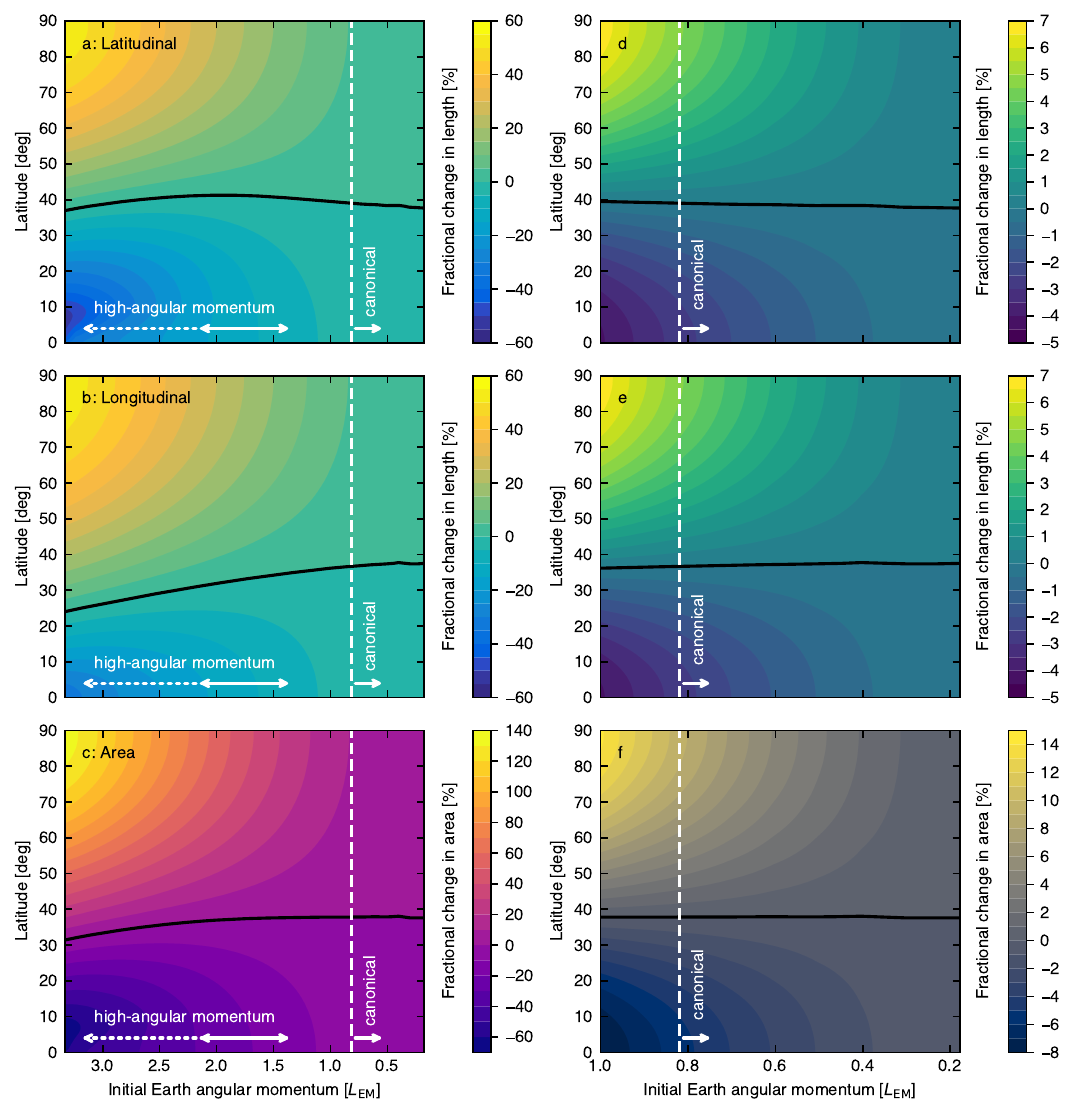}
\caption{Caption on next page.}
\end{figure}
\addtocounter{figure}{-1}
\begin{figure}
\centering
\caption{\add{The cumulative change in Earth's surface between an initial angular momentum and the present day, demonstrating that} Earth's surface \change{was}{could have been} substantially deformed during lunar tidal evolution as Earth's angular momentum was reduced and its rotation rate slowed. Shown are the cumulative fractional changes in local latitudinal (a, d) and longitudinal (b, e) length, and local surface area (c, f) as a function of latitude due to Earth losing angular momentum from a given initial angular momentum (x-axis) to that of the present day Earth (0.18~$L_{\rm EM}$). The left column (a--c) shows the results for the full range of possible initial terrestrial angular momenta after Moon formation and the right column (d--f) focuses on the range of initial angular momenta in the canonical model, where the angular momentum of the Earth-Moon system is constant over time. Solid black lines indicate the latitude of zero cumulative deformation. White dashed lines indicate the angular momentum of Earth if the Moon formed at the Roche limit ($\sim$~$2.9 R_{\rm Earth}$) in the canonical model, the maximum initial angular momentum of Earth in this scenario. Double-headed white arrows indicate the range of angular momenta explored in tidal (solid) and impact (dashed) studies of high-angular momentum Moon-formation models.}
\label{fig:total_change}
\end{figure}

Although the total magnitude of change is controlled by the initial angular momentum, the strain rate, and thus the driver for tectonic processes, is dependent on the exact tidal evolution history. In the following sections I will present calculations of the strain rate on early Earth in the canonical and high-angular momentum Moon-formation models. To compare deformation rates in different tidal evolution scenarios to those accommodated by modern subduction zones \cite{VanKeken2011} and mid-ocean ridges \cite{Gale2013}, I integrate the instantaneous strain as if deformation in the polar and equatorial regions is accommodated independently, with the latitudinal or longitudinal direction accommodated by a single feature in each case. In this approximation, the longitudinal deformation would be accommodated by a tectonic feature along a line of longitude, with the deformation integrated over minor circles of constant latitude. Maximum convergence would be at the equator and maximum extension at a variable latitude in the polar regions. Latitudinal deformation would be accommodated separately in the equatorial and polar regions along a line of latitude with the deformation integrated between the latitudes of zero deformation in the northern and southern hemispheres (between roughly 20 and 40$^{\circ}$, Figure~\ref{fig:total_change}).

\subsubsection{The canonical scenario}
\label{sec:results:recession:canonical}

\begin{figure}
\centering
\includegraphics[scale=0.98]{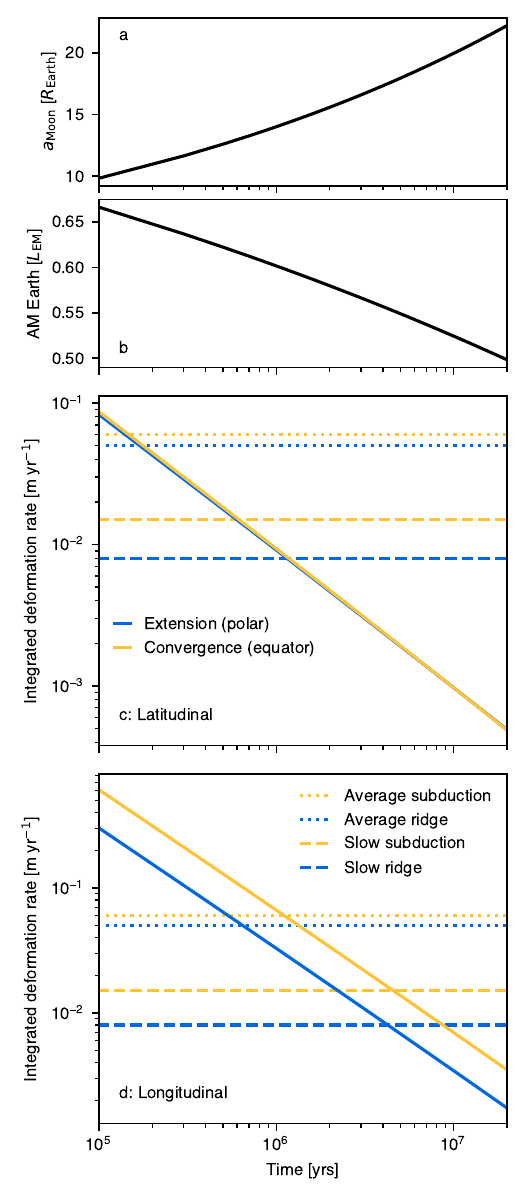}
\caption{Caption on next page.}
\end{figure}
\addtocounter{figure}{-1}
\begin{figure}
\centering
\caption{\change{Earth's crust was deformed substantially during lunar tidal recession.}{The orbital evolution and deformation of Earth's crust in a canonical scenario, demonstrating that} even in the canonical scenario, crustal deformation rates early in lunar tidal evolution \change{are}{could have been} greater than those accommodated by tectonic features on Earth today. Shown are the lunar semi-major axis (a), the angular momentum of Earth (b), and the maximum integrated convergence (yellow) and extension (blue) rates in the latitudinal (c) and longitudinal (d) directions, in the canonical scenario calculated using a simple tidal evolution model. In this scenario, the Moon is assumed to have formed at the Roche limit ($2.9$~$R_{\rm Earth}$), and both the tidal Love number ($k_{2,{\rm Earth}}$) and tidal quality factor ($Q_{\rm Earth}$) of Earth were constant and set to those of the present-day solid Earth, 0.3 and 100 respectively \cite{Yoder1995,Ray2012}. For comparison, horizontal lines show the convergence rate of an average subduction zone ($\sim$60~mm~yr$^{-1}$, dotted yellow lines) and a slow subduction zone (South Antilles, $\sim$15~mm~yr$^{-1}$, dashed yellow lines), and the spreading rate of an average mid-ocean ridge ($\sim$65~mm~yr$^{-1}$, dotted blue lines) and a slow spreading mid-ocean ridge (the Gakkel ridge, $\sim$8~mm~yr$^{-1}$, dashed blue lines) \cite{Gale2013,VanKeken2011}. The latitudinal variation in strain rate for this example is shown in Figure~\ref{sup:fig:canonical_example}, and Movie~\ref{sup:movie:canonical_example} contains an animated version of elements of this figure showing the change in shape and deformation rate during lunar tidal evolution.
} 
\label{fig:canonical_PTcomp}
\end{figure}

\begin{figure}
\centering
\includegraphics[height=0.95\textheight]{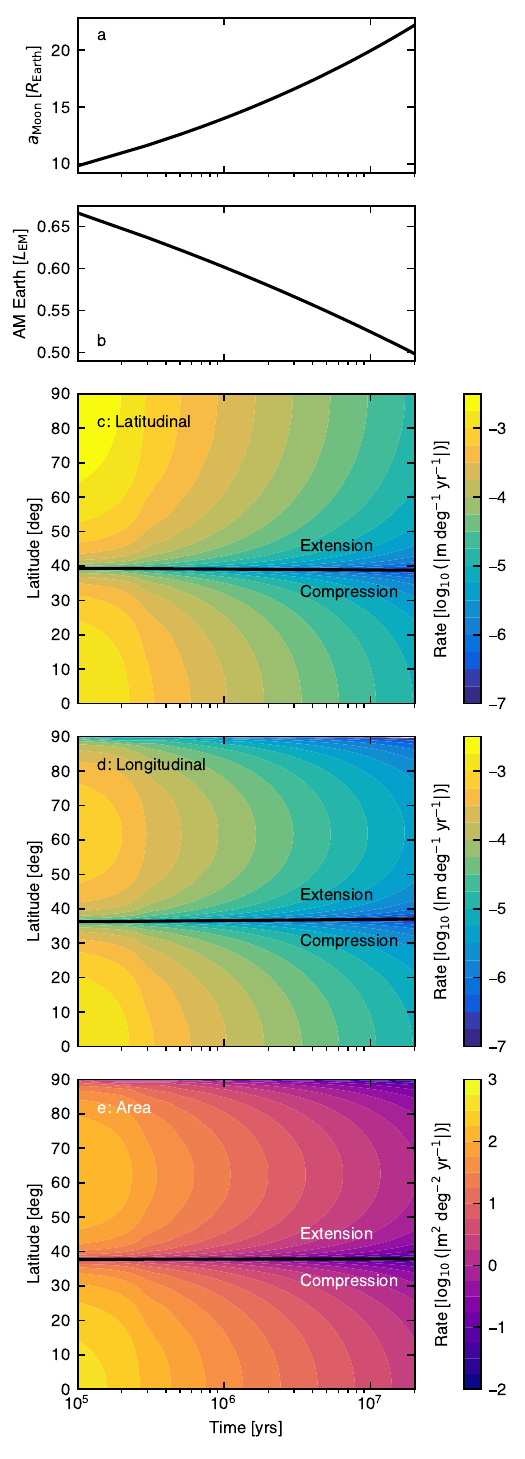}
\caption{Caption on next page.}
\end{figure}
\addtocounter{figure}{-1}
\begin{figure}
\centering
\caption{\add{The local rate of change of the surface as a function of time and latitude in the canonical scenario, demonstrating that }the rate of crustal deformation varies significantly over the planet. Shown are the lunar semi-major axis (a), angular momentum of Earth (b), rate of change of local length in the latitudinal (c) and longitudinal (d) directions, and the rate of change in local surface area (e) as a function of time in the canonical scenario calculated using a simple tidal evolution model. For reference, the average deformation rate across the Himalayan convergence zone at the present day is approximately 2~mm~deg$^{-1}$~yr$^{-1}$ \cite{Ader2012}, and the speed of mid-ocean ridges and subduction zones on present-day Earth are on the order $10^{-5}$--$10^{-4}$~m~deg$^{-1}$~yr$^{-1}$ when averaged over the circumference of Earth \cite{Gale2013,VanKeken2011}. In this scenario, the Moon is assumed to have formed at the Roche limit ($2.9$~$R_{\rm Earth}$), and the tidal Love number ($k_{2,{\rm Earth}}$) and tidal quality factor ($Q_{\rm Earth}$) of Earth were constant and set to those of the present-day solid Earth, 0.3 and 100 respectively \cite{Yoder1995,Ray2012}. The black lines in c--e indicate the latitude of zero instantaneous strain. The maximum integrated deformation rates in this scenario are shown in Figure~\ref{fig:canonical_PTcomp}, and Movie~\ref{sup:movie:canonical_example} contains an animated version of elements of this figure showing the change in shape and deformation rate during lunar tidal evolution.
}
\label{sup:fig:canonical_example}
\end{figure}

In the canonical scenario, rates of deformation of Earth's crust during lunar tidal recession are greater than or equivalent to those accommodated by tectonic features on Earth today for the first few million years. The \add{time evolution of the} maximum integrated strain rate for an example canonical scenario is shown in Figure~\ref{fig:canonical_PTcomp} with the latitudinal variation in strain rate shown in Figure~\ref{sup:fig:canonical_example}. The highest deformation rates occur when the Moon is close to Earth as the lunar recession rate is strongly dependent on semi-major axis with $\textrm{d}a/\textrm{d}t \propto a^{-11/2}$ (\add{Equations}~\ref{sup:eqn:adot}\add{ and }\ref{sup:eqn:n}\remove{Murray and Dermott, 1999}). The longitudinal convergence rates at the equator are greater than that accommodated by average ($\sim$60~mm~yr$^{-1}$, dotted yellow line in Figure~\ref{fig:canonical_PTcomp}) and slow (e.g., South Antilles, $\sim$15~mm~yr$^{-1}$, yellow dotted line) subduction zones for over one and approximately five million years, respectively. In the polar regions, the longitudinal extension rates are greater than those of average ($\sim$50~mm~yr$^{-1}$, blue dotted line) and slow-spreading (e.g., Gakkel, $\sim$10~mm~yr$^{-1}$, blue dashed line) ridges for half a million years and more than four million years, respectively. The integrated latitudinal deformation rates are lower, but the extension in the polar regions is still greater than that accommodated by a slow spreading ridge for over a million years. There continues to be a small amount of strain for much longer, hundreds of millions of years, but the rates are likely much less than from other sources such as mantle convection (see discussion in Section~\ref{sec:discussion:formation}).

\remove{Even in the canonical scenario, the rate of tidal recession of the Moon is highly uncertain due to the difficulty in calculating the tidal properties of the early Earth and Moon (Henning \& Hurford, 2014; Zahnle et al., 2015; Korenaga, 2025a, 2025b, 2025c). Furthermore, the radius at which the Moon forms can vary from the Roche limit ($\sim2.9$~$R_{\rm Earth}$) to several $R_{\rm Earth}$, introducing additional uncertainty in the orbital history (Kokubo et
al., 2000; Salmon \& Canup, 2012). However, although the rate of tidal recession can vary substantially,} I have found that, in the canonical scenario the rate of crustal deformation as a function of time is relatively insensitive to Earth's tidal properties and the initial semi-major axis of the Moon (for further details see \ref{sup:sec:tidal_model:dependence}). The amount by which Earth's shape changes due to a given change in \add{Earth's }angular momentum \add{($\mathrm{d}\delta l (\theta)/\mathrm{d}L_{\rm Earth}$)} decreases as the magnitude of Earth's angular momentum decreases\add{ (Figure}~\ref{sup:fig:beta}\add{b)}. If Earth is less dissipative, its angular momentum decreases more slowly, but this is compensated for by the fact that the absolute value of its angular momentum is higher at any given time, so a smaller change in angular momentum leads to greater surface deformation. In the simplest case, where Earth's tidal properties are constant, these competing effects almost cancel and the deformation rate is very weakly dependent on the tidal properties of Earth (Figure~\ref{sup:fig:k2Q_dependence}). In more realistic scenarios, where Earth's ability to dissipate energy increases over time \add{as the magma ocean freezes} \cite{Zahnle2015}, crustal deformation rates around the end of the magma ocean and primary crust formation can be several times higher than if the tidal parameters are held constant (Figure~\ref{sup:fig:increasing_k2Q}). 

Although the deformation rate is relatively insensitive to the tidal parameters of Earth, Earth's rotation rate at any given time can vary substantially. However, for reasonable values of the tidal parameters, Earth remains rapidly rotating and significantly oblate for at least a few millions years. Over the same time, the pressure in the crust increases by a few percent (Figure~\ref{sup:fig:gravity}).

\subsubsection{High-angular momentum scenarios}

The rates of crustal deformation driven by lunar tidal recession are substantially higher and more prolonged in high-angular momentum scenarios than in the canonical scenario. The greater total change in angular momentum leads to a much greater cumulative change in surface area (Figure~\ref{fig:total_change}a--c) and, furthermore, the complex dynamics by which angular momentum is transferred away from the Earth-Moon system can lead to periods of rapid change in Earth's angular momentum  \cite{Cuk2012,Wisdom2015,Cuk2016,Tian2020,Rufu2020,Ward2020,Cuk2021}.

Figure~\ref{fig:CukFig7_PTcomp} shows the maximum integrated crustal deformation rates for an example \add{Laplace plane instability} high-angular momentum tidal evolution history proposed by \citeA{Cuk2016,Cuk2021} with the latitudinal variation in strain rate shown in Figure~\ref{sup:fig:Cuk_example}. \remove{In this example, the Earth-Moon system had an initial angular momentum of 2.13~$L_{\rm EM}$ and an obliquity of 62$^{\circ}$ and the tidal properties of the Earth and Moon were constant with $k_{2,{\rm Earth}}/Q_{\rm Earth}=1\times10^{-2}$ and $k_{2,{\rm Moon}}/Q_{\rm Moon}=2\times10^{-2}$. The full orbital evolution of the Earth-Moon system in this example is presented in Figure~7 of \`{C}uk et al. (2021).} The results for an additional example of the \change{same fundamental}{Laplace plane instability} scenario, but with different assumed tidal parameters, are shown in Figure~\ref{sup:fig:CukFig3_PTcomp}. \remove{In this scenario, as the Moon-forming impact leaves Earth with a high obliquity, angular momentum is transferred from the Earth-Moon system to the Earth-Sun system during the so-called Laplace plane transition.} Throughout the first 10s~Myr of lunar evolution, the crustal deformation rates are roughly an order of magnitude greater than in the canonical scenario. The equatorial longitudinal convergence rates are greater than those accommodated by typical subduction zones \add{($\sim$60~mm~yr$^{-1}$)} for the first 30~Myr and greater than those accommodated by slow subduction zones \add{(e.g., South Antilles, $\sim$15~mm~yr$^{-1}$) }for over 90~Myr (Figure~\ref{fig:CukFig7_PTcomp}). A relatively common feature of high-angular momentum scenarios are periods of rapid angular momentum change, in this example marked by sudden decreases (`crashes') in the lunar semi-major axis\add{ when instabilities associated with the Laplace plane transition cause the lunar eccentricity to spike} \cite{tian_investigating_2026}. These rapid changes in angular momentum lead to roughly million year long spikes in deformation rate on the order of mm~deg$^{-1}$~yr$^{-1}$, comparable to the average deformation rate across the Himalayan convergence zone \cite<$\sim$2~mm~deg$^{-1}$~yr$^{-1}$,>{Ader2012}. However, unlike the Himalayas, the deformation on early Earth would be occurring at this rate around the entire equator.

\begin{figure}
\centering
\includegraphics[scale=0.98]{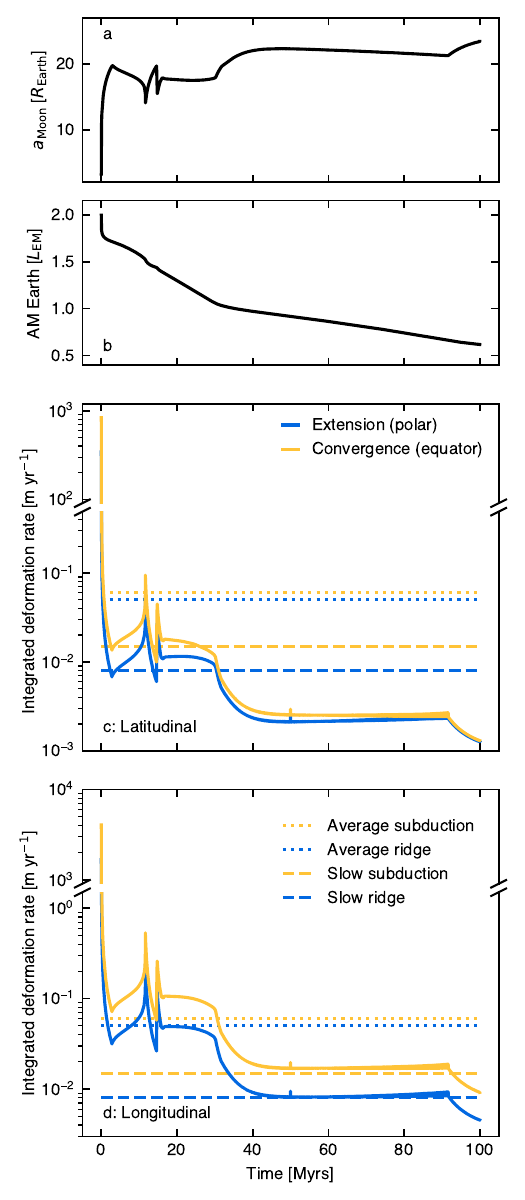}
\caption{Caption on next page.}
\end{figure}
\addtocounter{figure}{-1}
\begin{figure}
\caption{\add{The orbital evolution and deformation rate of Earth's crust in an example high-angular momentum scenario, demonstrating that} the deformation rates in high-angular momentum models are substantially larger than in the canonical model. Presented is the time dependent crustal deformation for the tidal evolution scenario shown in Figure~7 of \citeA{Cuk2021} where Earth initially has a large obliquity causing angular momentum to be transferred away from the Earth-Moon system during the Laplace plane transition. In this particular example, the Earth-Moon system has a total initial angular momentum of 2.13~$L_{\rm EM}$ and Earth begins with an obliquity of 62$^{\circ}$. The tidal properties of the Earth and Moon were constant with $k_{2,{\rm Earth}}/Q_{\rm Earth}=1\times10^{-2}$ and $k_{2,{\rm Moon}}/Q_{\rm Moon}=2\times10^{-2}$. Panels, lines and colors are the same as those in Figure~\ref{fig:canonical_PTcomp}. \add{The strain rates shown for present-day tectonic features are the same as in Figure}~\ref{fig:CukFig7_PTcomp} \cite{Gale2013,VanKeken2011}\add{.} Note that, unlike in Figure~\ref{fig:canonical_PTcomp}, the time axis is linear rather than logarithmic. 
The latitudinal variation in strain rate for this example is shown in Figure~\ref{sup:fig:Cuk_example}, and Movie~\ref{sup:movie:Cuk_example} contains an animated version of elements of this figure showing the change in shape and deformation rate during lunar tidal evolution.
} 
\label{fig:CukFig7_PTcomp}
\end{figure}

\begin{figure}
\centering
\includegraphics[height=0.95\textheight]{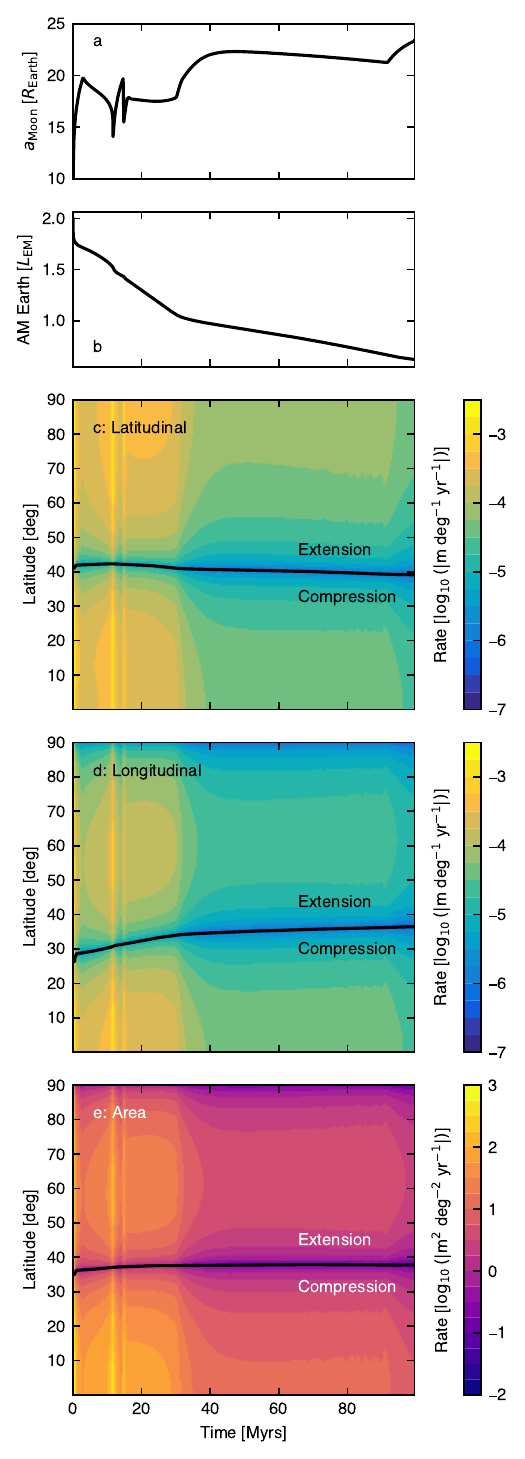}
\caption{Caption on next page.}
\end{figure}
\addtocounter{figure}{-1}
\begin{figure}
\centering
\caption{\add{The local rate of change of the surface as a function of time and latitude in an example high-angular momentum scenario, demonstrating that} crustal deformation rates during lunar tidal evolution are substantially larger in high-angular momentum Moon-formation scenarios. Shown are the semi-major axis of the Moon as a function of time (a), the angular momentum of Earth (b), the rate of change of local length in the latitudinal (c) and longitudinal (d) directions, and the rate of change in local surface area (e) for an example high-angular moment tidal evolution model \cite{Cuk2021}. The detailed lunar orbital evolution for this example is shown in Figure~7 of \citeA{Cuk2021}. In this scenario, the Earth-Moon system has a total initial angular momentum of 2.13~$L_{\rm EM}$ and Earth begins with an obliquity of 62$^{\circ}$. The color bars in c-e are the same as in Figure~\ref{sup:fig:canonical_example}. The black lines in c--e indicate the latitude of zero instantaneous strain. The maximum integrated crustal deformation rates are shown in Figure~\ref{fig:CukFig7_PTcomp}, and Movie~\ref{sup:movie:Cuk_example} contains an animated version of elements of this figure showing the change in shape and deformation rate during lunar tidal evolution.
}
\label{sup:fig:Cuk_example}
\end{figure}

The crustal deformation rates vary considerably between different high-angular momentum models, depending on the initial angular momentum, the mode of angular momentum transfer, and the tidal parameters of the Earth and Moon. This is well demonstrated using examples of another proposed mechanism for transferring angular momentum away the Earth-Moon system: a limit cycle associated with the evection resonance \cite{Wisdom2015}. Figure~\ref{fig:Rufu_comparison} shows a comparison of the maximum integrated longitudinal convergence rates in three different `successful' simulations of the limit cycle mechanism from the work of \citeA{Rufu2020}. Each example starts with the same initial value for the Earth-Moon system angular momentum (2.2~$L_{\rm EM}$) and reproduces the current angular momentum of the system, but assumes a different set of fixed tidal parameters ($A$ parameters of 5, 500, and $10^4$, and tidal time constants of $4.9\times 10^6$, $2.0\times 10^7$, and $1.5\times 10^8$~s, respectively). The large differences in tidal parameters between each example lead to angular momentum transfer timescales, and hence crustal deformation rates, that vary by two orders of magnitude. If the tidal dissipation in Earth just after Moon formation was large and comparable to that in the Moon (as in example 1 in Figure~\ref{fig:Rufu_comparison}) the angular momentum transfer, and hence the greatest change in shape, would have happened very early, \change{probably}{potentially} before the crust formed, and the deformation rates after the first million years or so would have been similar to that in the canonical scenario. However, if the early Earth was less dissipative (as in examples 2 and 3 in Figure~\ref{fig:Rufu_comparison}) crustal deformation rates many times those accommodated in typical mid-ocean ridges and subduction zones on Earth today could have been sustained for millions to tens of millions of years. These examples illustrate the extent to which the tidal response of the early Earth and Moon dictate the rate and duration of crustal deformation.

\begin{figure}
\centering
\includegraphics[]{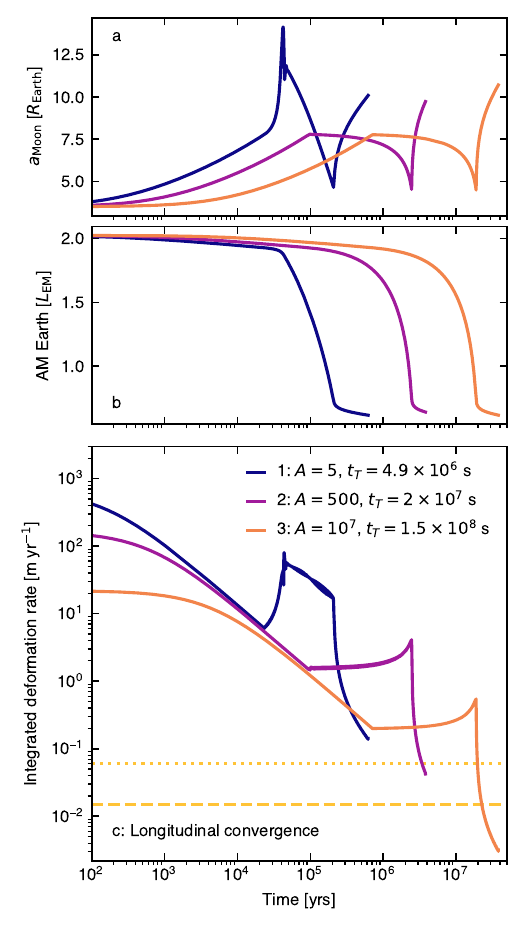}
\caption{Caption on next page.}
\end{figure}
\addtocounter{figure}{-1}
\begin{figure}
\centering
\caption{\add{A comparison of the orbital evolution and deformation rate of Earth's crust in high-angular momentum tidal evolution models with different tidal parameters, demonstrating that, }in high-angular momentum scenarios, the rate and timescale of surface deformation depends significantly on the tidal response of the Earth and Moon. Shown are the lunar semi-major axis (a), the angular momentum of Earth (b), and the maximum integrated convergence rates in the longitudinal direction (c) for three different simulations (colors) of the scenario in which angular momentum is transferred away from the Earth-Moon system when the system is captured in a limit cycle associated with the evection resonance \cite{Rufu2020,Wisdom2015}. Each example starts with the same initial value for the Earth-Moon system angular momentum (2.2~$L_{\rm EM}$) and reproduces the current angular momentum of the system, but assumes a different set of fixed parameters that control the tidal response of the system (the tidal parameter, $A$, and the tidal time constant, $t_T$; for definitions see \remove{Section~2}\ref{sup:sec:tidal_params}). Each simulation ends after the system has exited the limit cycle and resumed a more standard orbital evolution similar to the canonical scenario (Figure~\ref{fig:canonical_PTcomp}). The horizontal lines in c show the convergence rate of an average subduction zone (dotted) and a slow subduction zone (dashed), as in Figure~\ref{fig:canonical_PTcomp}. Figures~\ref{sup:fig:RufuA5_PTcomp}--\ref{sup:fig:RufuA10k_PTcomp} show the full latitudinal and longitudinal extension and convergence rates for each of these simulations and Movies~\ref{sup:movie:RufuA5}--\ref{sup:movie:RufuA10k} contain animated versions of elements of those figures. The eccentricity evolution of the scenario labeled as example 1 is shown in Figure~5 of \citeA{Rufu2020}.}
\label{fig:Rufu_comparison}
\end{figure}

As in the canonical scenario, the rotation rate of Earth at any given time in high-angular momentum models is highly variable depending on the angular momentum transfer mechanism and the tidal properties of the Earth and Moon. However, due to the fundamental need for time for any given mechanism to act to transfer angular momentum away from the Earth-Moon system, the Earth remains at a higher rotation rate for longer than in an equivalent canonical scenario. In many cases, the Earth can remain rotating with a period of only a few hours for tens of millions of years. As a consequence, the \add{gravity field would have remained significantly perturbed, and }internal pressures in the crust would have remained 10s\% lower than the present day\add{,} for tens of millions of years (Figure~\ref{sup:fig:gravity}\remove{b}). \remove{, increasing in response to decreases in Earth's angular momentum. Such pressure increases would drive phase changes in the crust. In particular, if the crust was significantly massive (more than $\sim3$ times the mass of the present-day crust), portions of the crust could have gone through the amphibolite to eclogite transition ($\sim1$--2~GPa, grey band in Figure~3b).}

\section{Discussion}
\label{sec:discussion}

In this section, I discuss the implications of the rapid and evolving rotation rate of Earth on its crust and surface environment. Given the considerable debate as to the viability of different angular momentum transfer mechanisms \cite{Cuk2012,Wisdom2015,Cuk2016,Ward2020,Rufu2020,Tian2020,Cuk2021} and the uncertainty in the tidal parameters of the early Earth and Moon \cite{Zahnle2015,Korenaga2025_tidal_3_compaction}, particular significance should not be placed on any one tidal evolution scenario. However, \add{I will assume, for the purposes of this discussion, that crust formation was rapid relative to tidal recession such that Earth's first crust}\remove{if, as expected (Elkins-Tanton, 2008; Lebrun et al., 2013; Zahnle et al., 2015; Korenaga, 2023), Earth's crust formed within a couple of million years of the Moon-forming impact, the Earth's first crust would have} formed on a rapidly rotating, flattened Earth. \remove{In addition, }Subsequent changes in Earth's shape caused by lunar tidal recession would then have provided a driver for crustal deformation and tectonic activity on early Earth. \remove{(Figure 10). These two previously unrecognized phenomena are active, although to greatly varying extents, in all possible lunar tidal evolution scenarios. }

\add{This assumption is made on the premiss that the timescale for solidification of the magma ocean (and therefore formation of the crust) is shorter than (or at least comparable to) the timescale for tidal evolution. In most magma ocean solidification models the crust forms within at most a couple of Myrs of the start of the simulation }\cite{Elkins-Tanton2008,Lebrun2013,Zahnle2015,Korenaga2023_rapid_solidification,Salvador2017,Hamano2013}\add{. This result is sensitive to the choice of atmospheric parameters and the planetary volatile abundance and partitioning of said budget between reservoirs, and some models that assume very volatile-rich atmospheres or certain atmospheric models can extend the magma ocean lifetime out to 10~Myr }\cite<e.g.,>{Salvador2017,Hamano2013}\add{. Meanwhile, the timescales over which there is significant surface strain for the tidal evolution simulations considered in this study range from millions to tens of millions of years, with a strong dependence on the assumed tidal parameters. Taking the models timings at face value, one might conclude that the crust formed very early on in many high-angular momentum models but could (on the longer end of magma ocean solidification timescales) have formed once the surface strains induced by changes in rotation rate were already low in the canonical model. However, there is likely an effective offset between the start of the magma ocean and tidal evolution simulations. Most tidal models assume that the Earth and the Moon have tidal properties similar to those of solidified planets, and so, in effect, the tidal simulations do not even begin until the surface of the magma ocean reached the rheological transition and the crust formed (see Section}~\ref{sup:sec:tidal_model}\add{). It is therefore possible that the full tidal evolution calculated in tidal models occurred while the Earth had a solid crust. For example, }\citeA{Korenaga2023_rapid_solidification}\add{ suggested that Earth's surface would solidify before the Moon reached $\sim7$--9 Earth radii away from the Earth. Observational constraints on the timescale of orbital evolution are limited but, recently, it has been argued, on the basis of the need for a large tidal heating event to explain the ages of lunar samples, that the Moon did not undergo the Laplace plane transition until $\sim150$~Myr after the formation of the Moon }\cite{nimmo_tidally_2024,tian_investigating_2026}\add{, by which time the Earth almost certainly had a crust. Conclusively determining the relative timescales of crustal formation and lunar tidal recession will likely require significant extra work, including coupling of tidal and thermal evolution models }\cite{Zahnle2015,Korenaga2025_tidal_1_inertia_e,Korenaga2025_tidal_2_atmo,Korenaga2025_tidal_3_compaction,Korenaga2023_rapid_solidification}\add{ and this uncertainty should be born in mind when considering the implications of our results. Nevertheless, based on the existing magma ocean and tidal evolution models, I consider that it is likely that Earth formed its first crust while rapidly rotating and substantially oblate, and that that crust witnessed the effects of changes of shape during lunar tidal recession. }

At present, we lack tools capable of making quantitative predictions of geophysical processes on the highly oblate and evolving surface of the early Earth. For the purposes of this exploratory study, I therefore present first-order calculations and qualitative descriptions of key processes that may have governed the surface environment and subsequent evolution of early Earth. \add{It is important to note that these previously unrecognized phenomena would have been active, although to greatly varying extents, in all possible lunar tidal evolution scenarios. }

\subsection{Formation of crust on a rapidly-rotating Earth}
\label{sec:discussion:formation}

\change{Previous work has modelled condensation of the post-impact body and}{The first (at least comparably) long-lived crust on Earth was formed during the }
solidification of the molten Earth (often referred to as the magma ocean) after the Moon-forming impact. \add{When the liquid fraction at the surface of the magma ocean fell below about 40\%, the solid crystals become locked together, and the viscosity of the material increases by many orders of magnitude} \cite{abe_thermal_1997,Solomatov2000,Solomatov1993}\add{. At this point, melt percolated upwards and erupted at the surface forming a compositionally distinct crust and leaving behind a depleted mantle. Previous work on this period of Earth's evolution has assumed}\remove{assuming} that the internal pressures, effective gravity etc. of early Earth were the same as \change{in}{at} the present day \cite{Elkins-Tanton2008,Lebrun2013,Zahnle2015,Solomatov2015,Miyazaki2019a,Miyazaki2022_heterogenious_mantle,Boukare2015,boukare_solidification_2025}. However, \add{if the tidal recession of the Moon was slow enough,  the first crust on Earth would have formed while Earth was still rapidly rotating and substantially oblate. In this case,} due to its rapid rotation, Earth's geophysical structure \remove{ during solidification of the magma ocean}(e.g., internal pressure, gravitational field) would have been, in fact, very different from that of the Earth today \cite{Lock2017,Lock2019pressure,Lock2020}. \remove{Significantly, as is evident from the tidal evolution scenarios presented here, the tidal recession of the Moon is slow enough that when the magma ocean solidified, and hence the first crust of Earth formed, Earth was still rapidly rotating and substantially oblate (see also Korenaga, 2023).} 
\add{I now turn to consider how formation in this altered geophysical environment could have affected the properties of Earth's first crust}. 

\remove{If the blanketing effect of the early atmosphere was low enough, the surface temperature could fall below the liquidus of the molten surface while much of the mantle was still liquid. If this was the case, the first solid surface of any kind might have been formed by quenching of the surface layer, as seen on lava lakes on Earth today. This can also be thought of in terms of a crust forming in the upper thermal boundary layer of the mantle/magma ocean (e.g., Solomatov, 2015). The thickness of the crust would be set by thermal conduction and so would be similar around the surface of the planet. However, unlike in a lava lake, the quenched crust would not be buttressed by the solid sides of the lake and would be highly unstable (Elkins-Tanton, 2005, 2008). Here I consider the nature of a more long-term stable crust that replaced any initial quenched crust.

The properties of this first crust are highly uncertain and depend on the relative timescales of crystal settling and melt percolation/matrix compaction in the magma ocean, and that of magma ocean cooling (Solomatov, 2015). The multiphase dynamics and thermodynamics of the magma ocean are complex and not well understood at the high-pressure conditions in a terrestrial magma ocean (Solomatov \& Stevenson, 1993b, 1993c, 1993a; Abe, 1993; Stixrude et al., 2009; Fiquet et al., 2010; Solomatov, 2015; Boukar\`{e} et al., 2015; Miyazaki \& Korenaga, 2019a; Wolf \& Bower, 2018). I will therefore not attempt to argue for a particular magma ocean solidification or crust formation scenario, but consider a range of scenarios and the effect of Earth's rapid rotation on each. 

First, it is important to note that the magma ocean may have started to crystallize from the middle of the mantle, forming what is known as a basal magma ocean near the core mantle boundary  (Labrosse et al., 2007; Ballmer et al., 2017; Caracas et al., 2019;
Boukar\`{e} et al., 2025) separated from the upper mantle by a solid layer. However, the existence of a basal magma ocean is unlikely to have any direct effect on the formation of the first crust, besides potentially altering the heat flow in the magma ocean and so the timescales of solidification. Here I will thus only concern myself with the fraction of the mantle that is solidifying upwards towards the surface. 

If crystal settling and/or melt percolation is rapid, then the solid phase on the liquidus is continuously added to a cumulate layer at the bottom of the magma ocean and the magma ocean itself becomes increasingly enriched in incompatible elements. This limiting scenario is known as fractional crystallization, an analogue for many magma chambers in volcanic systems, and is the assumption made in many models of magma-ocean solidification (Elkins-Tanton, 2008; Boukar\`{e} et al., 2015). The result is a solid cumulate that is highly chemically stratified and likely unstable to density-driven overturn (Elkins-Tanton, 2008) and mixed continuously through solidification (Maurice et al., 2017; Boukar\`{e} et al., 2018; Ballmer et al., 2017; Miyazaki \& Korenaga, 2019b, 2022). In the opposite extreme, where crystal settling and melt percolation are both slow compared to the cooling time, no fractionation can occur and the solidified mantle has a homogenous composition with depth  (Solomatov \& Stevenson, 1993b).

In reality, the timescales for each of the key processes varies significantly between different stages of magma ocean solidification, and different places in the system (Solomatov, 2015). In particular, when the liquid fraction in a particular part of the magma ocean falls below about 40\%, the solid crystals become locked together, the material goes through a rapid rheological transition, and its viscosity increases by many orders of magnitude (Abe, 1997; Solomatov, 2000; Solomatov \& Stevenson, 1993b). When this rheological front reaches the surface marks an important transition in magma ocean evolution, in particular for crustal formation. 

While the surface of the magma ocean is still molten, the cooling timescale is dependent on the mass, composition and structure of Earth's atmosphere as the atmosphere is the limiting control on the transport of heat out of the mantle. Estimates for the timescale to reach solidification of the surface vary from $10^5$ to $10^6$~yrs (Lebrun et al., 2013; Miyazaki \& Korenaga, 2019b; Elkins-Tanton, 2008; Korenaga, 2023), largely depending on the structure, composition, and optical properties of the atmosphere. Once the surface of the magma ocean solidifies, heat transport is then limited by conduction across a thermal boundary layer at the surface, greatly increasing the cooling timescale and allowing more time for percolation of melts to the surface to form a crust.

When the rheological front reaches the surface, melt will percolate upwards and erupt at the surface forming a compositionally distinct crust and leaving behind a depleted mantle.}

The thickness of the crust \change{will depend}{depends} on the efficiency of melt extraction and the volume from which melt is extracted. On a rapidly rotating early Earth, the average lower gravity and pressure gradient \change{means}{would mean} that a larger volume of the mantle \change{will}{would have} still \change{be}{been} partially molten when the surface \change{reaches}{reached} the rheological transition than on a planet with the present-day Earth structure. The total volume of melt available to be extracted from the mantle to form the crust is therefore larger. Further, the available melt would be greater at the equator than at the poles due to the lower gravity and pressure gradient at the equator (Figure~\ref{sup:fig:gravity}) with the depth of melting ($D$) scaling as $D \sim1/g$, where $g$ is the gravitational acceleration. However, the timescale for melt extraction ($t_{\rm perc}$) would have been longer nearer the equator as a result of both the reduced buoyancy driver for separation of the melt and solid due to the lower gravity, and the greater depth over which melt must be extracted. 
\begin{equation}
t_{\rm perc} = \frac{D}{v_{\rm perc}} \; ,
\end{equation}
where
\begin{equation}
v_{\mathrm{perc}}=\frac{d^2 g \Delta \rho}{150 \eta} \frac{\phi^2}{1-\phi} \; ,
\end{equation}
$d$ is the grain size of the solid, $\Delta \rho$ is the density difference between the melt and solid, $\eta$ is the melt viscosity, and $\phi$ is melt fraction \cite{dullien_porous_2012,soo_fluid_1967}. The timescale for extraction of melt from the whole partially-molten layer therefore scales as
\begin{equation}
    t_{\rm perc}\sim \frac{1}{g^2} \; .
\end{equation}
It is important to note that the timescale to extract a given amount of melt scales as $1/g$. The time available to extract the melt from the partially molten layer can be approximated from the convective velocity ($v_{\rm conv}$) with the timescale for entraining the partial melt layer into the mantle given by
\begin{equation}
t_{\rm conv}=\frac{D}{v_{\rm conv}} \; .
\end{equation}
The convective velocity is dependent on the Rayleigh number ($\mathrm{Ra}$) and \change{so}{the regime of convection. In the simplest case of thermally-driven, isoviscous convection }
\begin{equation}
    v_{\rm conv}\sim \mathrm{Ra}^{\frac{2}{3}} \sim g^{\frac{2}{3}} \; 
    \label{eqn:vconv}
\end{equation}
\cite{turcotte2014}\add{. Stagnant-lid convection would alternatively give exponents from 0.5 to $\sim1.3$, depending on the rheology} \cite{korenaga_scaling_2009}\add{.} Note that due to the high viscosity of the solid mantle and \change{crystal-locked, partially molten region}{partially molten yet crystal-locked regions}  \cite{Solomatov2015}, the Coriolis force arising from Earth's rapid rotation has a negligible effect on convection, unlike in the fluid-dominated regions of the magma ocean \cite{Maas2019,Solomatov1993}. $t_{\rm conv}$ therefore scales as $t_{\rm conv}\sim g^{-5/3}$ \add{in the isoviscous case and with an exponent between -1.5 and $\sim-2.3$ in the case of stagnant-lid convection. In each case, $t_{\rm conv}$ scales}\remove{,} very similarly to the timescale for extracting melt from the entire partial molten volume. It has been convincingly argued in other works that the timescale for melt extraction from the partially-molten region on a body with the gravity of present-day Earth is shorter than the convective timescale and so melt extraction should be efficient \cite{Miyazaki2022_heterogenious_mantle,Solomatov2015}. Given the \add{similar }scaling of the timescales for melt extraction and convection, I conclude that \change{this}{it} is also likely the case \add{that melt extraction was efficient} on a rapidly-rotating Earth and so \change{a decreased gravity will likely result in a, potentially much, thicker average crust with thickness varying latitudinally (Figure~10a)}{the thickness of the crust would have varied latitudinally due to the varying gravitational field.}  

\begin{figure}
\centering
\includegraphics[]{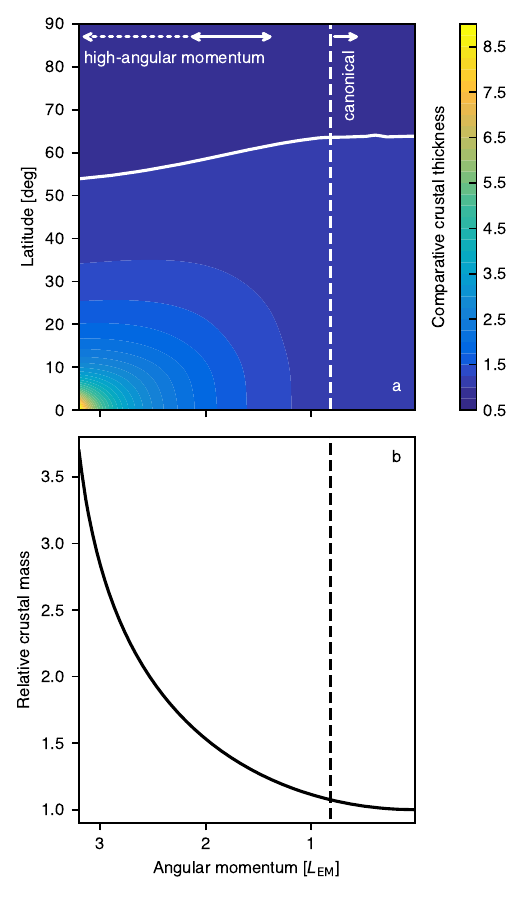}
\caption{\add{Caption on next page.}}
\end{figure}
\addtocounter{figure}{-1}
\begin{figure}
\centering
\caption{\add{Crustal thickness (a) and mass (b) on the early Earth as a function of the angular momentum at which the crust formed, demonstrating how rapid rotation can lead to the production of more voluminous and latitudinally-varying crust. The crustal thickness and mass are normalized to that which would be produced on a non-rotating Earth, following the arguments in Section}~\ref{sec:discussion:formation}\add{. The white contour in a shows the latitude at which the crust is the same thickness as in the non-rotating case. The white/black dashed lines indicates the angular momentum of Earth if the Moon formed at the Roche limit in the canonical scenario. Double-headed white arrows indicate the range of angular momenta explored in tidal (solid) and impact (dashed) studies of high-angular momentum Moon-formation models.}
}
\label{fig:crustal_D_V}
\end{figure}

\remove{The volume and chemistry of the melt extracted from the partially molten but crystal-locked layer depends, once again, on how the bulk of the magma ocean solidified. If the magma ocean froze (at least somewhat) fractionally, then the residue liquid would be enriched in elements more incompatible in the solid cumulate. This incompatible-enriched composition would then be inherited by the crystal-locked layer and that layer would likely be denser than the average mantle. In contrast, if the mantle froze homogeneously then the crystal-locked layer would have the same composition as the bulk mantle, and any melt extracted from that layer would be mafic or ultramafic, more similar to present-day oceanic crust.

If the crystal-locked layer (before or after melt extraction) or the melt that is extracted from it is denser than the mantle below it, Rayleigh-Taylor instabilities would rapidly cause it to overturn into the underlying mantle (Maurice et al., 2017; Boukar\`{e} et al., 2018; Ballmer et al., 2017; Miyazaki \& Korenaga, 2019b, 2022). The upwelling mantle that replaced the crust and/or crystal-locked mantle would melt due to decompression and melt would be extracted to form additional, or a replacement, crust. Similarly, crust production driven by upwellings in solid-state mantle convection, for example in environments akin to modern-day mid-ocean ridges or plumes  (e.g., Moore \& Webb, 2013), would also be governed by decompression melting. The composition of said crust would be very dependent on the mode of magma ocean solidification. If magma ocean solidification was homogenous the material that would be melting would have a composition similar to that of the bulk silicate Earth (BSE). If the freezing of the magma ocean had been fractional, then the average mantle would be very depleted leading to the production of a thin crust with the potential of smaller-scale pockets of more incompatible element-rich material left over from the overturn of the cumulate during magma ocean freezing (Miyazaki \& Korenaga, 2022). The thickness of crust produced by decompression melting during overturn or long-lived mantle convection would again be dependent on surface gravity, with the volume of crust produced by standard decompression melting models inversely proportional to the gravitational acceleration (Korenaga, 2006). }

\remove{In summary, in most realistic crust-formation scenarios, a thickness of crust formed on a rapidly rotating early Earth would be a function of local gravitational acceleration. the lateral variations in effective gravity and pressure would result in the crust being potentially many times thicker in the equatorial regions than near the poles (Figure~10a). Further, the crust could have been much more voluminous than previously assumed --- by a factor of several --- due to the lower average effective gravity (Figure~3), with large differences depending on the angular momentum of Earth when crust formation occurred. }

\add{Figure}~\ref{fig:crustal_D_V}\add{ shows how the thickness of crust at different latitudes depending on the angular momentum of Earth at the time the crust was formed (a), and the total mass of crust produced (b), if the crust was all formed at the same time. Following the arguments above, I assumed that the mass of melt extracted was not limited by percolation and so scales as $1/g$. Note that, while I have been primarily discussing crust formed by percolation of melt at the end of the magma ocean, the thickness of crust formed later by other processes, such as during overturn of an existing crust }\cite{Maurice2017,Boukare2018,Ballmer2017,Miyazaki2019a,Miyazaki2022_heterogenious_mantle} \add{or in environments akin to mantle plumes or spreading ridges} \cite<e.g.,>{Moore2013}\add{, would also depend on the local gravitational field and vary latitudinally. The exact thickness, volume and chemistry of the crust formed will depend on the potential temperature and whether the melt is being derived from a well-mixed mantle, or one that had been enriched (e.g., by fractional crystalization of the magma ocean) or depleted (e.g., by previous crust extraction). Given these uncertainties, the crustal thickness and mass shown in Figure}~\ref{fig:crustal_D_V}\add{ are referenced to that which would be produced on a non-rotating Earth.}

\add{In the canonical model, the latitudinal variation in crustal thickness would be modest with a variation from equator to pole of $\sim$15\%. However, if the angular momentum of Earth was much higher, the crustal thickness variation could be as much as a factor of ten. The higher potential temperature on early Earth would have led to more extensive melting than on the comparatively cold present-day Earth, the effect of which itself could have led to an increase in crustal thickness by a factor of three }\cite{Weller2019}\add{. Combining this effect with the effect of rapid rotation (Figure}~\ref{fig:crustal_D_V}\add{), the mass of crust produced on Earth in the highest angular momentum scenarios could be an order of magnitude larger than if the same process was acting on the present-day Earth. A latitudinally varying crust would modulate the heat flow out of the mantle and }a very large volume of melt extracted from the mantle during production of early crust \add{(Figure}~\ref{fig:crustal_D_V}\add{b)} could have a substantial impact on the long-term geochemical and geophysical evolution of both the mantle and crust, potentially with observationally testable signatures (see Section~\ref{sec:discussion:ancient_reservoirs}).

\subsection{Response of Earth's earliest crust to changes in shape during lunar tidal evolution}
\label{sec:discussion:response}

Understanding the implications of Earth's changing shape during tidal evolution for the terrestrial crust is complicated by the fact that a lot of crustal material (defined by its composition) was not within the lithosphere (Figure~\ref{fig:cartoon}). That is, the high heat flow and potentially thick compositionally-distinct crust of this early epoch led to a reversal of the modern situation in which the crust is always embedded within the thermal boundary layer. Once the crust formed, the timescale for the thermal boundary layer to extend to the base of the crust was on the order of 10$^6$--10$^8$~yrs, depending on crustal thickness. Assuming that the crust was bouyant compared to the depleted mantle and/or bulk mantle, there was therefore a substantial fraction of the crust that was within the asthenosphere, and responding viscously/ductilely to forcing, during the 10$^6$--10$^7$~yr period of substantial crustal strain. I will refer to this section of the crust as the asthenospheric crust and the overlying rheologically-solid crust as the lithospheric crust. A similar distinction must be made when considering the response of the mantle depleted to form the crust, if such is long-lived. The portions of the crust that lie in the lithosphere or asthenosphere will depend on the time relative to the extraction of the crust and hence the thickness of the thermal boundary layer, but also on the forcing timescale \cite{lau_toward_2020}. For example, in cases of rapid deformation, less of the crust would respond viscously and more of the crust would effectively be in the lithosphere. It is important to consider the different responses and fates of the lithospheric and asthenospheric crust during the early phases of lunar tidal recession.

Initially, the compositionally-defined crust would have been thicker in the equatorial regions than near the poles, due to the large variation in surface gravity (Figure~\ref{sup:fig:gravity} and Section~\ref{sec:discussion:formation}). As the shape and gravity field of Earth evolved, the equipotential surfaces became more spherical and any asthenospheric crust would have been redistributed latitudinally. The pressure in the crust also increased as the angular momentum of Earth decreased \cite{Lock2019pressure}, potentially causing the crust to undergo phase changes. Figure~\ref{sup:fig:gravity}b shows the pressure at the base of crusts of different masses as a function of angular momentum, assuming that the crust flowed rapidly enough to always lie on an equipotential. If the crust was massive enough (greater than approximately three times the present day mass of crust \add{as is likely in high-angular momentum scenarios, Section}~\ref{sec:discussion:formation}), garnet may have been stabilized at the bottom of what was originally buoyant asthenospheric crust during the course of lunar tidal evolution \add{(Figure}~\ref{sup:fig:gravity}\add{b)}. The higher density of the garnet-bearing eclogite would lead to layers of the crust becoming denser than the mantle and delaminating, falling lower into the mantle and potentially contributing to ancient chemical and physical heterogeneity in the mantle (see Section~\ref{sec:discussion:ancient_reservoirs}). Reduction in the mass of the crust due to delamination would also strongly influence the subsequent evolution of the planet. 

Deformation of the lithospheric crust led to thickening of the lithosphere in the equator and thinning at the poles. The mechanisms by which strain was accommodated is uncertain, and would have depended strongly on the rheology of the warm early crust at the frequency of forcing, and on the presence of any additional stresses or drivers of deformation. \remove{The mode of mantle convection that regulated Earth's evolution following the Moon-forming impact is, however, uncertain. The higher temperatures of the Hadean mantle would have reduced its viscosity and so potentially also the strength of coupling between the lithosphere and the convecting mantle (Korenaga, 2021a; Al Asad \& Lau,
2024). In the most extreme case, the Earth could have been in a stagnant lid regime (O’Neill \& Zhang, 2019) with the surface consisting of a single rigid plate with communication from the mantle to the surface through intraplate volcanism (Moore \& Webb, 2013). More likely, the early Earth was in an intermediate sluggish lid mode  where the lithosphere would have been still loosely coupled to the mantle (Al Asad \& Lau, 2024), including the possibility of the planet being in a ``plutonic squishy lid'' mode (Van Kranendonk, 2010; Gerya et al., 2015; Fischer \& Gerya, 2016; Rozel et al., 2017). It has even been argued that for the first tens to hundreds of millions of years after the Moon-forming impact the Earth was temporarily in a mobile lid regime with strong coupling between the mantle and the plates causing rapid plate motions, more akin to modern-day plate tectonics (Korenaga, 2021a; Miyazaki \& Korenaga, 2022). }
In the end member case where Earth was in a stagnant lid regime \add{early in its history }\cite{ONeil2019_early_crust_book,Moore2013}, some insight into the response of the crust to changes in shape can perhaps be gained from previous studies of other bodies in the solar system \cite{JayMelosh1977,Dobrovolskis1982,Matsuyama2008,Matsuyama2010,Rhoden2012}. These studies made assumptions that are not applicable to early Earth, such as small deformation from a sphere and/or constant density bodies, and therefore are most applicable to the canonical scenario where the planet is closer to spherical. Given the comparatively large volume contraction of early Earth in response to the increased pressure as its rotation rate slowed, the work of \citeA{JayMelosh1977} would suggest the deformation of a brittle, previously undamaged, lithosphere would be accommodated by thrust faulting in the equatorial regions with strike-slip and then normal faults at higher latitudes. However, even in this end-member scenario, strain could still be localized along pre-existing faults or damage zones and may even have been accommodated elastically in strong crustal zones.

If deformation of the terrestrial crust was occurring within a background of a more active tectonic regime\add{, with various sluggish (including ``plutonic squishy'') lid }\cite{al_asad_coupled_2024,VanKranendonk2010,Gerya2015,fischer_early_2016,Rozel2017}\add{ and mobile lid }\cite{korenaga_hadean_2021,Miyazaki2022_heterogenious_mantle}\add{regimes having been proposed for early Earth,}
the deformation could be accommodated by increased motion at existing plate boundaries, including altering the sense of motion of said boundaries. In both cases, sufficient forcing could lead to breaking of existing plates and the formation of new plate boundaries. Other forces, such as planetary impacts \cite<e.g.,>{ONeill2017} or solid tides raised by the Moon, could also play a role in the localization and sense of deformation of the crust. Future studies that incorporate the evolving shape of Earth into mantle convection models will be required to determine which mode of accommodating the deformation was dominanant on early Earth.
 
\subsection{An increase in lithological diversity early in Earth's history}
\label{sec:discussion:litho}

\add{It is generally assumed that the first crust on Earth formed by direct melting of the bulk mantle and was (ultra)mafic }\cite<see e.g.,>{Taylor_hadean_2008}\add{. However, evidence from ancient zircons and the oldest rocks strongly suggests that Earth had its first enriched (i.e., more felsic) crust shortly after the Moon-forming impact }\cite{Wilde2001,Harrison2008,Hopkins2008,Harrison2009,Wang2018,Harrison2017,Harrison2008}\add{. There have been several mechanisms proposed to produce more felsic crust on early Earth }\cite<e.g.,>{Miyazaki2022_heterogenious_mantle,Marchi2014}\add{, but the geophysical and geochemical processes that led to divergence of felsic and mafic crust  are still hotly debated.}

\remove{As discussed above, the petrology, thickness, and stability of Earth's first crust depends strongly on the relative rates of cooling, crystal settling, melt extraction, and fluid dynamical instabilities in the cooling magma ocean (Solomatov \& Stevenson, 1993b; Solomatov, 2015; Solomatov \& Stevenson, 1993a, 1993c; Korenaga, 2021a; Miyazaki \& Korenaga, 2022).  It is often assumed that the first crust was formed by direct melting of the bulk mantle and was (ultra)mafic (see e.g., Taylor \& McLennan, 2008) but scenarios have been proposed where mantle heterogeneities produced in the process of solidifying the mantle (e.g., Miyazaki \& Korenaga, 2022) or burial by impact ejecta (e.g., Marchi et al., 2014) could have led quickly to lithological variation. Even if the first terrestrial crust was largely mafic, there is evidence to suggest it did not stay that way for long. The properties of zircons (very robust crystals that are all that remain of the earliest rocks) as old as 4.4~Ga have been interpreted as evidence that Earth hosted its first felsic rocks within approximately 100~Myrs of the Moon-forming giant impact (Wilde et al., 2001; Harrison et al., 2008; Hopkins et al., 2008; Harrison, 2009). The Hf isotope systematics of a larger suite of Hadean and Archean zircons also indicate the emergence of an enriched (i.e., more felsic) reservoir shortly after the Moon-forming impact (Wang \& Wilde, 2018; Harrison et al., 2017, 2008). The geophysical and geochemical processes that led to this divergence of felsic and mafic crust are hotly debated. 


The Moon-forming giant impact occurred late in accretion (Lock, Bermingham, et
al., 2020) and Earth likely had a substantial fraction of its current volatile budget after the impact (Halliday, 2013), despite the potential for volatile loss due to the impact itself (Genda \& Abe, 2003, 2005; Lock \& Stewart, 2024; Kegerreis et al., 2020; Roche et al., 2025). After the Moon-forming giant impact some fraction of volatiles would have been left in the atmosphere as the magma ocean rained out of the hot silicate vapor of the post-impact body (Caracas \& Stewart, 2023; Stewart et al., 2018). Combined with subsequent degassing during solidification of the magma ocean (Elkins-Tanton, 2008; Bower et al., 2022) and/or during mantle convection (Korenaga, 2021a)), the post-impact Earth likely had a significant atmosphere, probably dominated by carbon species (Lupu et al., 2014; Sossi et al., 2020; Schaefer \& Fegley, 2017). Early on there might also have been a substantial mass of steam in the atmosphere, but the impactor flux and geothermal heat would have been low enough that most of the water in the atmosphere would have condensed to form an ocean within a few million years of the impact (Abe \& Matsui, 1988; Zahnle et al., 2007). Observationally, oxygen isotope measurements of some of the oldest surviving zircons provide evidence for a water cycle on Earth within at most two hundred million years of the Moon-forming impact (Mojzsis et al., 2001). Earth would therefore have had an ocean and atmosphere while the crust was being deformed during lunar tidal recession.} 

\add{Tectonics driven by the changing shape of Earth during lunar tidal evolution could aid in the production of lithologically diversity. Early Earth likely had an atmosphere dominated by carbon species }\cite{Lupu2014,Sossi2020,Schaefer2017} \add{and a water ocean would have condensed very shortly after the surface of the magma ocean became rheologically solid }\cite{Abe1988,Zahnle2007,Mojzsis2001}\add{.} Interaction with the atmosphere and surface water cycle would have hydrated and/or weathered the upper crust. Yielding of the lithospheric crust in the equatorial regions could have transported such material to depth and, hence, to higher pressures and temperatures. Water and carbon both lower the melting point of silicates \cite<e.g.,>{Dasgupta2013}, and transport of surface material to greater depth could partially melt crustal material and produce evolved magmas. Separation and freezing of these magmas could produce more felsic rocks.\remove{, which may explain the existence of a lithologically diverse crust within 10s~Myrs of the Moon-forming impact as evidenced in the early zircon and rock records (Harrison, 2009; Carlson et al., 2019; Nebel et al., 2014; Wilde et al., 2001; Harrison et al., 2008; Hopkins et al., 2008).} 

Lithological diversity would have been further increased by decompression melting due to extension and faulting in the polar regions, a process somewhat analogous to that occurring at mid-ocean ridges today. However, in contrast to mid-ocean ridges, the material that would be drawn up and melted could not just be from the upper mantle, but also asthenospheric crust or depleted mantle from the extraction of the initial crust. Remelting of crustal composition material would result in the production of secondary melts, without the need for the addition of volatiles, but with a petrology quite different to melting of hydrated or weathered surface material. 

\add{These processes, driven by the changing shape of Earth, may explain the existence of a lithologically diverse crust within 10s~Myrs of the Moon-forming impact as evidenced in the early zircon and rock records }\cite{Harrison2009,Carlson2019,Nebel2014,Wilde2001,Harrison2008,Hopkins2008}\add{. } 

\subsection{Implications for the surface environment of early Earth}
\label{sec:discussion:environment}

Little is currently known about the surface environment of early Earth, but constraining the conditions during this early epoch is necessary, among other things, to determine the potential for the emergence and survival of life \cite{kitadai_origins_2018,Arndt2012,Sutherland2016}. It is therefore important to consider the implications of the processes described in this work on the surface environment of early Earth. 

As well as the lithology of the crust (Section~\ref{sec:discussion:litho}\add{)}, of particular interest is when sub-aerial topography (i.e., dry land) emerged \cite{Korenaga2021_was_there_land}, and so when pathways for the emergence of life requiring sub-aerial environments become feasible \cite<e.g.,>{Sutherland2016}. Despite the potential for significant crustal convergence in the equatorial regions during lunar tidal recession, the thin lithosphere during this early epoch may not have been able to rheologically sustain any part of the crust having substantial, long-lived topography \cite<although dynamic topography may allow parts of the surface to be transiently sub-aerial;>{guimond_blue_2022}. However, secondary melting of the primary crust (Section~\ref{sec:discussion:litho}) and separation of those melts from the residue could have produced lower density volumes of crust. These sections of the crust would have gained higher elevations in isostatic equilibrium, in the same way as the continents do today, potentially large enough to be sub-aerial (Figure~\ref{fig:cartoon}c). Further modeling is required to quantify the density of any evolved crust that could be produced and so evaluate the potential for the generation of sub-aerial topography.

An early period of tectonic activity would change the chemistry of the atmosphere and oceans. For example, extension in the polar regions would have driven upwelling, melting, and thus outgassing of a portion of the mantle and crust. Weathering associated with even short-lived topographic uplift could have enriched the early oceans in Ca and led to the sequestration of atmospheric carbon in abiotic carbonates. Burial of these carbonates, along with extraction of carbon into weathered surface material, would have reduced the mass of the early, dense, carbon-dominated atmosphere and helped produce a more habitable surface environment \cite{Zahnle2007}. The ocean chemistry would also have been altered by a potentially massive increase in hydrothermal activity associated with tectonic activity, facilitating the transfer of many elements into the oceans and making \change{them}{these elements} more readily available for potential organisms. Modification of the chemistry of both the atmosphere and ocean is a significant outcome of lunar tidal recession and should be explored further.

\subsection{Formation of ancient geochemical reservoirs}
\label{sec:discussion:ancient_reservoirs}

The formation, delamination and/or melting of crust in the manner described in Sections~\ref{sec:discussion:formation} and \ref{sec:discussion:response} could have created distinct isotopic and chemical reservoirs early in Earth's history. For example, the large volume of melting to form the earliest crust followed by secondary melting of sections of the crust would have concentrated heat-producing elements (U, K, Th) into the crust (particularly more evolved crust) very early, modifying the heat sources that drive mantle convection. Delamination of the lower asthenospheric crust would have transported enriched material deep into the Earth, leaving the mantle \add{(at least temporarily) }as a complementary depleted reservoir. The lifetime of the heterogeneities produced by recycling of the earliest crust, and its potential to be isolated from the convecting mantle long-term, would depend on the lengthscale of the instability and the vigor of mantle convection in this early epoch. As the delaminated crust formed early, within the lifetime of $^{146}$Sm, the enriched and depleted reservoirs would have distinct short-lived radiogenic Nd isotope anomalies, potentially consistent with Sm-Nd isotopic systematics of Archean rocks \cite<e.g.,>{Carlson2019}. Delamination of volumes of early crust could also play a role in\change{the origin of }{ generating mantle heterogeneity, for example they could contribute a fraction of the mass of} the present-day Large Low Shear Velocity Provinces and Ultra Low Velocity Zones \cite<LLSVPs \& ULVZs, volumes of material at the base of the mantle that have anomalous seismic velocities;>{mcnamara_review_2019}\add{ and the anomalous geochemical signatures of ocean-island basalts thought to be associated with them } \cite<e.g.,>{parai_noble_2025,Carlson2015,Lock2020ssr}\add{.}

\section{Conclusion}
\label{sec:conclusion}

\begin{figure}
\centering
\includegraphics[width=\textwidth]{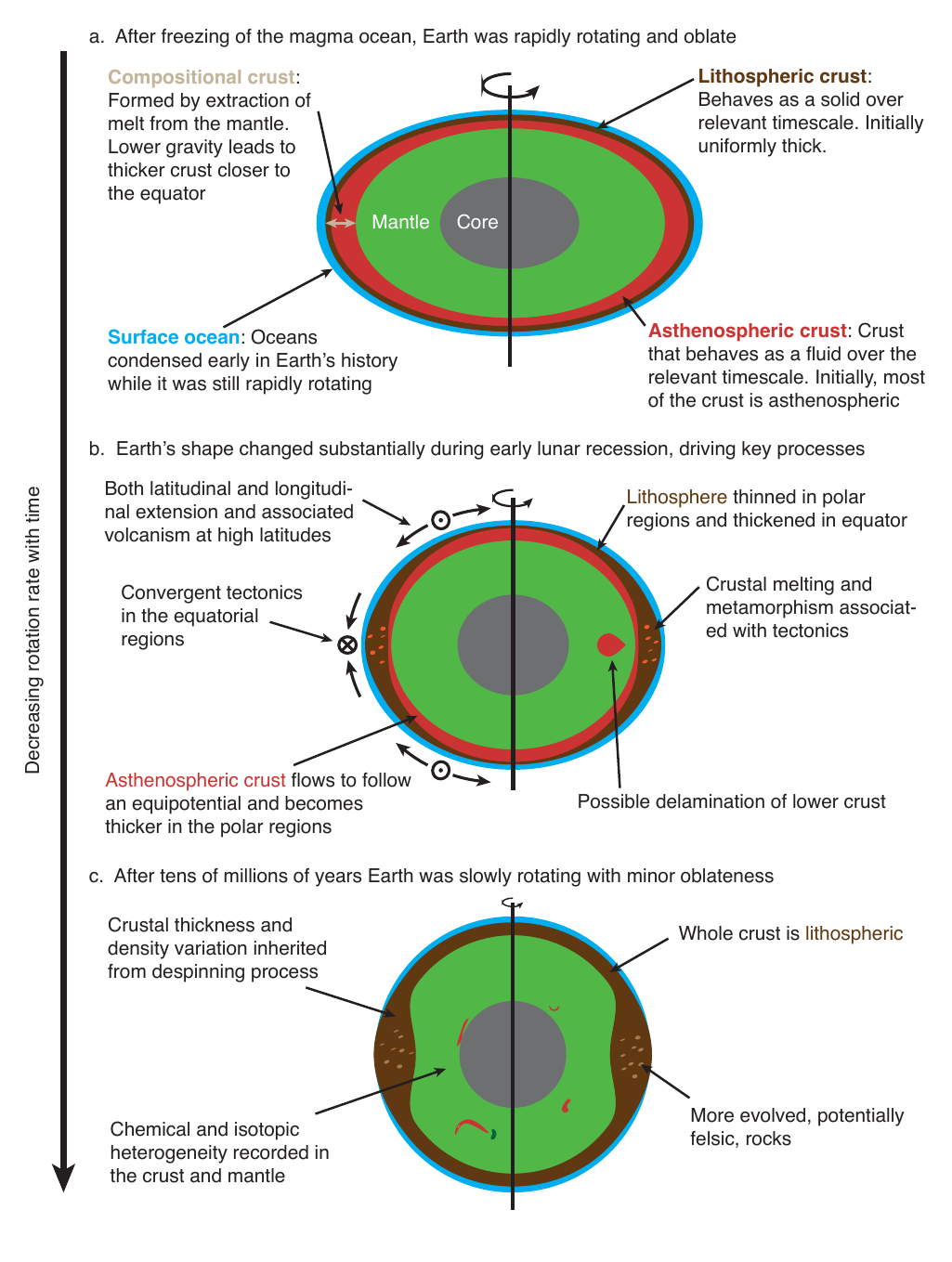}
\caption{Caption on next page.}
\end{figure}
\addtocounter{figure}{-1}
\begin{figure}
\caption{A conceptual illustration of Earth's evolution during the early period of lunar tidal recession. The depth of the crust and ocean are exaggerated, but the oblateness of Earth is to scale for the aftermath of a high-angular momentum Moon-forming impact (Figure~\ref{fig:shape}). When the first crust on Earth formed the planet was rotating rapidly (a). The internal pressures were lower than at the present day and the effective gravity was stronger at lower latitudes. A large fraction of the crust could have been asthenospheric and the compositional crust (i.e., that with a composition distinct from the mantle) thickened towards the equator. During lunar tidal recession Earth's shape evolved, driving crustal deformation (b). The lithospheric crust was extended in the polar regions and compressed in the equatorial regions, both latitudinally and longitudinally. Compressional tectonics in the equatorial regions led to a thickening of the lithospheric crust, potentially melting crustal material and producing more felsic rocks. The asthenospheric crust flowed to accommodate the change in the gravitational field. The increase in pressure in the crust could have led to delamination of a fraction of the asthenospheric crust. As the Moon moved farther from Earth the shape of Earth approached spherical (c). Tectonics driven by the changing shape of Earth during lunar tidal recession could have left Earth with a heterogeneous crust, both in thickness and composition.  }
\label{fig:cartoon}
\end{figure}

The early rapid rotation of Earth, and its subsequent decrease during lunar tidal recession, had significant consequences for Earth's early crust and surface environment. Earth was rotating with only a few hour period when its surface first solidified, and the first crust on Earth would have likely been thicker at the equator than at the poles by as much as a factor of ten \add{in the highest angular momentum models} (Figure\add{s}~\ref{fig:crustal_D_V}\add{a and }\ref{fig:cartoon}a). The change in shape of Earth as its rotation rate slowed during lunar tidal recession provided a significant driver for deformation of this crust \add{and/or modulated the response of the crust to other forces }(Figure~\ref{fig:cartoon}b). Extension in the polar regions and compression nearer the equator would have been at rates equivalent to the deformation accommodated by plate boundaries on Earth today for millions to hundreds of millions of years after Moon formation, with short-lived peaks\add{, in high angular momentum models, }when deformation was equivalent to that accommodated in the Himalayas today but around the entire equator.

How the crust responded to this change in shape depended on its rheology and any simultaneous drivers of tectonics. Deformation of the crust could have led to increased outgassing of the mantle, production of evolved melts with and without the incorporation of hydrated/weathered surface material, and development of sub-aerial topography (Figure~\ref{fig:cartoon}c). Delamination or subduction of parts of the crust would also have introduced geochemical domains into the mantle that have distinct chemistry and radiogenic isotope signatures. These processes could explain key parts of the early zircon and rock record, and geochemical tracers of silicate differentiation. 

Primary crust formation while Earth was rapidly-rotating and tectonics driven by the changing shape of Earth during lunar tidal recession are natural outcomes of Moon formation from a giant impact, with the magnitude dependent on the initial angular momentum of the Earth-Moon system and its subsequent orbital evolution. \add{Although the effect of these processes would be strongest in a high-angular momentum Moon-formation scenario }\cite{Cuk2012,Canup2012,Lock2018moon}\add{ the latitudinal variations in crustal properties and the addition of a background stress field could still play a role in Earth's subsequent evolution in a canonical Moon-formation scenario.} Future work on this earliest period of Earth's evolution must consider the effect of rotation on the dynamics and geochemistry of Earth. Such an undertaking will likely require advances in geodynamic codes to simultaneously track the thermal and orbital evolution of the Earth-Moon system \cite{Korenaga2025_tidal_1_inertia_e,Korenaga2025_tidal_2_atmo,Korenaga2025_tidal_3_compaction,Zahnle2015,Korenaga2023_rapid_solidification} and account for the deformed and evolving shape of the body on the gravitational field, internal pressures etc. More advanced and coupled geochemical and geophysical models will allow us to determine the extent to which rotation dictated the early evolution of Earth\add{, potentially test different Moon-formation models,} and so learn more about this shrouded period in Earth's history.

\appendix

\section{Dependence of deformation rate on tidal properties and initial conditions in the canonical model}
\label{sup:sec:tidal_model:dependence}

In Section~\ref{sec:results:recession:canonical}, I noted that the rate of deformation in the canonical scenario is relatively insensitive to Earth's tidal parameters. Below I explain the physics behind this somewhat surprising result from my simple tidal evolution model for the canonical scenario (Section~\ref{sup:sec:tidal_model}).

Figure~\ref{sup:fig:k2Q_dependence} shows the lunar semi-major axis and equatorial deformation rate at different times (colored lines) as a function of $k_{2,{\rm Earth}}/Q_{\rm Earth}$ (x-axis) and the initial semi-major axis of the Moon ($a_0$, line styles). $k_{2,{\rm Earth}}/Q_{\rm Earth}$ and the total angular momentum of the system are assumed to be constant in each case. In cases where the Moon starts close to Earth (e.g., $a_0=2.9$~$R_{\rm Earth}$, solid lines) the deformation rate is relatively insensitive to $k_{2,{\rm Earth}}/Q_{\rm Earth}$ over the times relevant for crustal deformation, varying by about a factor of two over five orders of magnitude in $k_{2,{\rm Earth}}/Q_{\rm Earth}$. For cases in which the Moon starts farther from Earth (e.g., $a_0=10$~$R_{\rm Earth}$, dotted lines) the deformation rate at earlier times, when the Moon has not moved appreciably outwards from its initial position, can be substantially lower. At later times, the deformation rates for the high and low $a_0$ cases converge.

\begin{figure}
\centering
\includegraphics[height=0.95\textheight]{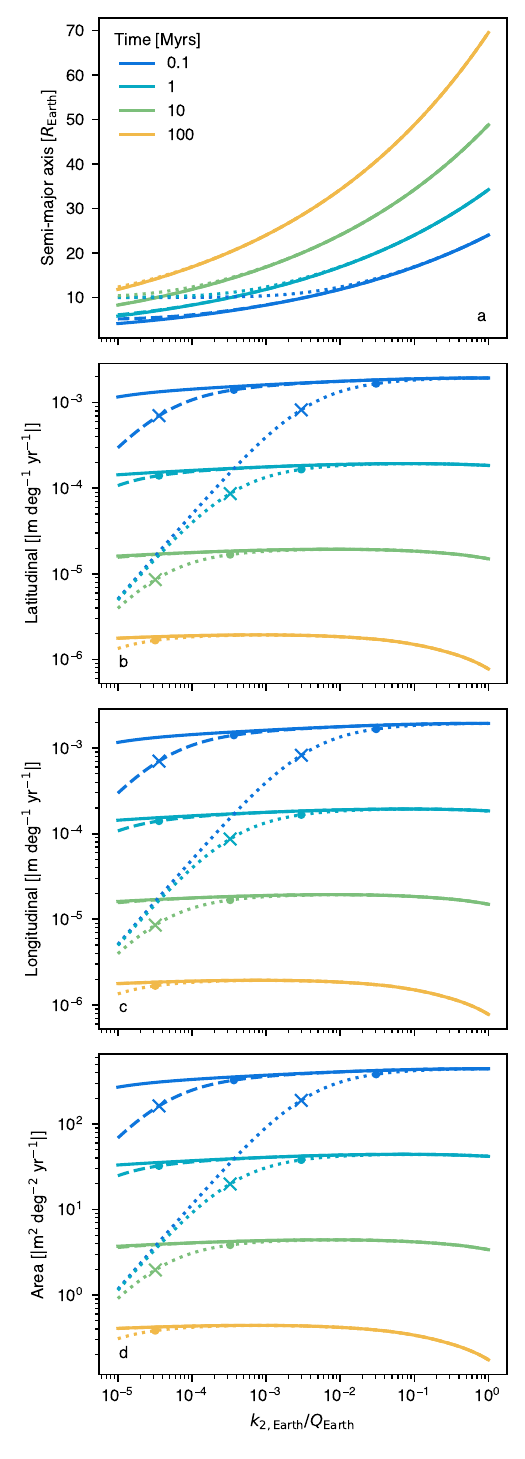}
\caption{Caption on next page.}
\end{figure}
\addtocounter{figure}{-1}
\begin{figure}
\centering
\caption{\add{Orbital evolution and crustal deformation as a function of tidal parameters in the canonical scenario, illustrating that the }crustal deformation rates \remove{in the canonical scenario }are relatively insensitive to Earth's tidal properties and the radius at which the Moon formed. Shown are the semi-major axis (a), rate of change of local length in the latitudinal (b) and longitudinal (c) directions, and the rate of change in local surface area (d) at varying times (colors) for canonical tidal models with different constant $k_{2,{\rm Earth}}/Q_{\rm Earth}$ (x-axis) and initial positions of the Moon (line styles). Values for the initial semi-major axis of the Moon of $2.9$~$R_{\rm Earth}$ (solid lines), $5$~$R_{\rm Earth}$ (dashed lines) and $10$~$R_{\rm Earth}$ (dotted lines) are presented. Deformation rates are shown for the equator. The crosses and dots in panels b--d show the point at which the time exceeds $t_{\rm tran}$ and $10$~$t_{\rm tran}$, respectively, for each time and $a_0$. $t_{\rm tran}$ is the transition between the early and late time regimes where the dependence of the deformation rate on $k_{2,{\rm Earth}}/Q_{\rm Earth}$ changes. In most cases, $t_{\rm tran}$ is exceeded before the time for all $k_{2,{\rm Earth}}/Q_{\rm Earth}$ shown, and so the transition is not marked. At early times and large $a_0$, the deformation rate increases linearly with $k_{2,{\rm Earth}}/Q_{\rm Earth}$. At later times, and for smaller $a_0$, the deformation rate is only weakly dependent on $k_{2,{\rm Earth}}/Q_{\rm Earth}$.
}
\label{sup:fig:k2Q_dependence}
\end{figure}

The behavior of the system, including the result that crustal deformation rates are relatively insensitive to tidal properties at later times, can be understood using simple scaling relationships. I will consider two temporal limits. First, in the late $t$ limit, where
\begin{equation}
\label{sup:eqn:tlimit}
    t \gg \frac{2}{39} \frac{Q_{\rm Earth}}{k_{2,{\rm Earth}}} \frac{M_{\rm Earth}}{M_{\rm Moon}\sqrt{G(M_{\rm Earth}+M_{\rm Moon})}} \left( \frac{1}{R_{\rm Earth}} \right )^5  a_0^{13/2} \, ,
\end{equation}
the first term in the brackets in Equation~\ref{sup:eqn:a_constk2Q} (the expression for the semi-major axis as a function of time) is dominant and the semi-major axis of the lunar orbit at time $t$ scales as 
\begin{equation}
\label{sup:eqn:ascaling}
    a \propto \left ( \frac{k_{2,{\rm Earth}}}{Q_{\rm Earth}} \right )^{2/13} t^{2/13} \, .
\end{equation}
It follows, using Equation~\ref{sup:eqn:dLdt_a}, that the rate of change of angular momentum with time scales as
\begin{equation}
\label{sup:eqn:dLdtscaling}
    \frac{\mathrm{d}L_{\rm Earth}}{\mathrm{d}t} \propto \left ( \frac{k_{2,{\rm Earth}}}{Q_{\rm Earth}} \right )^{1/13} t^{-12/13} \, .
\end{equation}
The scaling of $L_{\rm Earth}$ with the tidal parameters and time is complicated by the fact that both the total angular momentum of the system and the angular momentum of the Moon are of comparable magnitude for most of tidal evolution (Equation~\ref{sup:eqn:L}). For clarity, I express the scaling of Equation~\ref{sup:eqn:L} as
\begin{equation}
\label{sup:eqn:Lscaling}
    L_{\rm Earth} \propto L_{\rm EM} - \xi \left ( \frac{k_{2,{\rm Earth}}}{Q_{\rm Earth}} \right )^{1/13} t^{1/13}   \, ,
\end{equation}
where 
\begin{equation}
\xi=  \left ( \frac{39}{2} \frac{R_{\rm Earth}^5}{M_{\rm Earth}} \right )^{1/13}  M_{\rm Moon}^{14/13} \left [ G ( M_{\rm Earth}+M_{\rm Moon}) \right ]^{7/13}
\end{equation}
is a constant for a given system.

In the canonical model the angular momentum of Earth can range from 0.18 - 0.82 $L_{\rm EM}$. In this angular momentum range, the rate of change of the local surface length and area at a given latitude as a function of angular momentum (as determined from the planetary structure calculations, Section~\ref{sup:sec:HERCULES}) is generally well described by a power law,
\begin{equation}
\label{sup:eqn:ddl_dL_pl}
    \frac{\mathrm{d}\delta l (\theta)}{\mathrm{d}L_{\rm Earth}} \propto L_{\rm Earth}^{\beta (\theta)} \,.
\end{equation}
In the equatorial and polar regions the dependence is roughly linear with $0.8<\beta < 1.2$ (Figure~\ref{sup:fig:beta}). The relationship breaks down at mid-latitudes (approximately 30$^{\circ}$--50$^{\circ}$) where the deformation regime is transitioning from compressional to extensional and is not well described by a single power law across the full range of angular momentum. The magnitude of deformation in these regions is small (Figure~\ref{fig:total_change}) and for the rest of this section I will use Equation~\ref{sup:eqn:ddl_dL_pl} as a simple descriptor of the dependence on the rate of change of the surface lengths with angular momentum.

\begin{figure}
\centering
\includegraphics{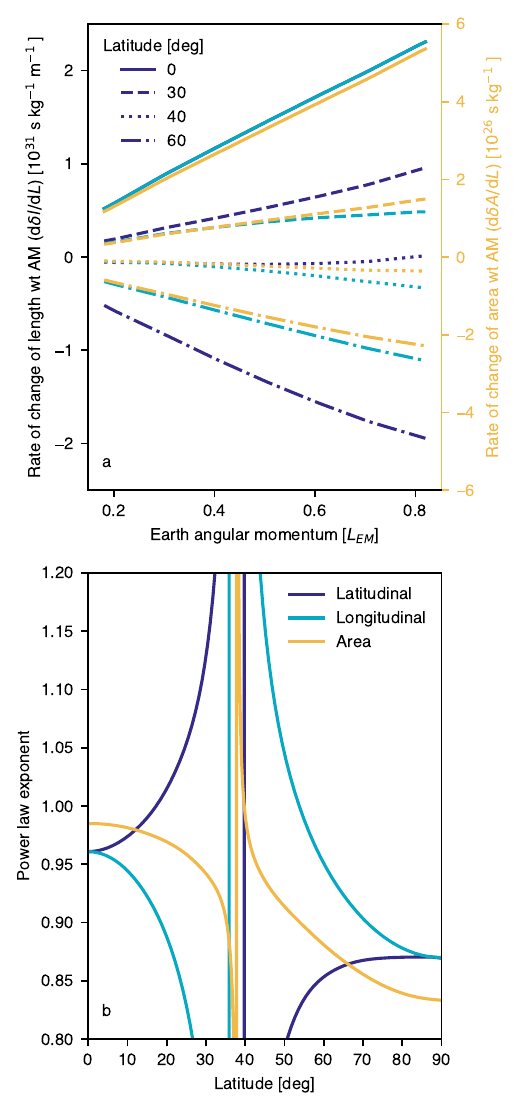}
\caption{Caption on next page.}
\end{figure}
\addtocounter{figure}{-1}
\begin{figure}
\caption{The rate of change of local surface length and area \add{as a function of angular momentum which} are close to \change{linearly dependent on angular momentum}{linear} in the equatorial and polar regions over the range of angular momenta relevant for the canonical scenario (0.18--0.82~$L_{\rm EM}$). Shown is \add{a: The rate of change of local length in the latitudinal (dark blue, left axis) and longitudinal (light blue, left axis) directions and change in the local surface area (yellow, right axis) as functions of angular momentum ($\mathrm{d}\delta l(\theta)/\mathrm{d}L_{\rm Earth}$) at different latitudes (line styles). b:} The fitted power law exponent for $\mathrm{d}\delta l(\theta)/\mathrm{d}L_{\rm Earth} \sim L_{\rm Earth}^{\beta}$ for change in local length in the latitudinal (dark blue) and longitudinal (light blue) directions, and change in the local surface area (yellow). At high and low latitudes the exponent is close to one (i.e., the relationship is linear). At mid-latitudes, where the deformation is small, $\mathrm{d}\delta l(\theta)/\mathrm{d}L_{\rm Earth}$ becomes highly non-linear as the deformation transitions from compressional to extensional.
}
\label{sup:fig:beta}
\end{figure}

Combining Equations~\ref{sup:eqn:dLdtscaling}, \ref{sup:eqn:Lscaling} and \ref{sup:eqn:ddl_dL_pl} 
\begin{eqnarray}
\label{sup:eqn:rate_scaling}
    \frac{\mathrm{d}\delta l(\theta)}{\mathrm{d}t} & = & \frac{\mathrm{d}\delta l(\theta)}{\mathrm{d}L_{\rm Earth}} \frac{\mathrm{d}L_{\rm Earth}}{\mathrm{d}t} \nonumber \\[10pt] 
    && \propto \left [ L_{\rm EM}  - \xi \left ( \frac{k_{2,{\rm Earth}}}{Q_{\rm Earth}} \right )^{1/13} t^{1/13} \right ]^{\beta} \left ( \frac{k_{2,{\rm Earth}}}{Q_{\rm Earth}} \right )^{1/13} t^{-12/13} \, .
\end{eqnarray}
As noted above, $\beta \sim$1 for most of Earth's surface, and taking $\beta =1$ gives
\begin{equation}
\label{sup:eqn:rate_scaling_alpha1}
    \frac{\mathrm{d}\delta l(\theta)}{\mathrm{d}t} \propto  L_{\rm EM} \left ( \frac{k_{2,{\rm Earth}}}{Q_{\rm Earth}} \right )^{1/13} t^{-12/13} - \xi \left ( \frac{k_{2,{\rm Earth}}}{Q_{\rm Earth}} \right )^{2/13} t^{-11/13}  \, .
\end{equation}
The dependence on the tidal parameters ($k_{2,{\rm Earth}}/Q_{\rm Earth}$) for both terms in Equation~\ref{sup:eqn:rate_scaling_alpha1} is weak. At early times, and lower $k_{2,{\rm Earth}}/Q_{\rm Earth}$, the first term dominates and the deformation rate increases slightly with increasing $k_{2,{\rm Earth}}/Q_{\rm Earth}$. At later times, and higher $k_{2,{\rm Earth}}/Q_{\rm Earth}$, the second term dominates and the deformation rate decreases with increasing $k_{2,{\rm Earth}}/Q_{\rm Earth}$. The transition between these two regimes occurs at lower $k_{2,{\rm Earth}}/Q_{\rm Earth}$ at later times due to the slightly stronger inverse dependence of the first term on $t$. 

Qualitatively, when $k_{2,{\rm Earth}}/Q_{\rm Earth}$ is higher, the Moon recedes more quickly from Earth and $\mathrm{d}L_{\rm Earth}/\mathrm{d}t$ is always higher at a given time (Equation~\ref{sup:eqn:dLdtscaling}). As a result, the angular momentum of Earth at a given time is always lower in higher $k_{2,{\rm Earth}}/Q_{\rm Earth}$ scenarios than in lower $k_{2,{\rm Earth}}/Q_{\rm Earth}$ scenarios. As Earth's angular momentum decreases, the change in shape associated with a given change in angular momentum also decreases. Therefore, even though Earth's angular momentum is decreasing faster in high $k_{2,{\rm Earth}}/Q_{\rm Earth}$ models, the effect that a given change in angular momentum has on the surface of Earth is less. The increase in $\mathrm{d}L_{\rm Earth}/\mathrm{d}t$ with $k_{2,{\rm Earth}}/Q_{\rm Earth}$ and the decrease in $\mathrm{d}\delta l(\theta)/\mathrm{d}t$ with $k_{2,{\rm Earth}}/Q_{\rm Earth}$ have similar dependencies, and so the net effect is that the deformation rate is relatively insensitive to $k_{2,{\rm Earth}}/Q_{\rm Earth}$ in the late time regime (as defined by Equation~\ref{sup:eqn:tlimit}).

Deviation from the relationships outlined above is seen at early times, and low $k_{2,{\rm Earth}}/Q_{\rm Earth}$, particularly for scenarios with larger $a_0$ (dotted and dashed lines in Figure~\ref{sup:fig:k2Q_dependence}). These deviations are due to the fact that the condition expressed in Equation~\ref{sup:eqn:tlimit} is not met. In the opposite, early $t$, limit
\begin{equation}
\label{sup:eqn:tlimit_low}
    t \ll \frac{2}{39} \frac{Q_{\rm Earth}}{k_{2,{\rm Earth}}} \frac{M_{\rm Earth}}{M_{\rm Moon}\sqrt{G(M_{\rm Earth}+M_{\rm Moon})}} \left( \frac{1}{R_{\rm Earth}} \right )^5  a_0^{13/2} \, ,
\end{equation}
the second term in the brackets in the expression for semi-major axis as a function of time (Equation~\ref{sup:eqn:a_constk2Q}) dominates, and the semi-major axis of the Moon is approximately constant. Intuitively, these are cases where the dissipation ($1/Q_{\rm Earth}$) is low enough that the lunar orbit has not evolved substantially by time $t$. Consequently, $L_{\rm Earth}$ is also approximately constant and the other relevant terms scale as:  
\begin{equation}
\label{sup:eqn:dadtscaling_low}
    \frac{\mathrm{d}a_{\rm Earth}}{\mathrm{d}t} \propto  \frac{k_{2,{\rm Earth}}}{Q_{\rm Earth}}  \, ,
\end{equation}
\begin{equation}
\label{sup:eqn:dLdtscaling_low}
    \frac{\mathrm{d}L_{\rm Earth}}{\mathrm{d}t} \propto  \frac{k_{2,{\rm Earth}}}{Q_{\rm Earth}}  \, ,
\end{equation}
and
\begin{equation}
\label{sup:eqn:ddldtscaling_low}
    \frac{\mathrm{d}\delta l (\theta)}{\mathrm{d}t} \propto  \frac{k_{2,{\rm Earth}}}{Q_{\rm Earth}}  \, .
\end{equation}
Therefore, at early times the deformation rate increases linearly with $k_{2,{\rm Earth}}/Q_{\rm Earth}$, as can be seen for some cases in Figure~\ref{sup:fig:k2Q_dependence}. There is no dependence of deformation rate on $t$ and so the deformation rate for cases with the same starting semi-major axis converges at low $k_{2,{\rm Earth}}/Q_{\rm Earth}$  (see e.g., dotted lines in Figure~\ref{sup:fig:k2Q_dependence}). Qualitatively, because Earth's angular momentum is not changing substantially over short timescales, the response of Earth's surface to a change in angular momentum is constant. Therefore, the dependence of the tidal recession rate on $k_{2,{\rm Earth}}/Q_{\rm Earth}$ dominates and higher $k_{2,{\rm Earth}}/Q_{\rm Earth}$ lead to greater deformation rates. 

\begin{figure}
\centering
\includegraphics{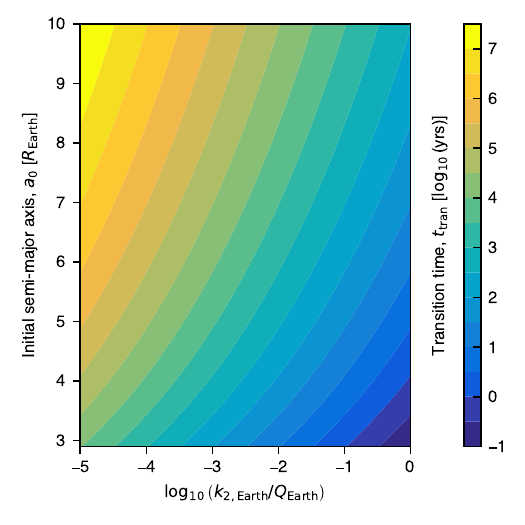}
\caption{The transition between the early and late time regimes for the dependence of deformation rates on $k_{2,{\rm Earth}}/Q_{\rm Earth}$, $t_{\rm tran}$, in the canonical scenario typically occurs early in tidal evolution. Shown are $t_{\rm tran}$ for varying $k_{2,{\rm Earth}}/Q_{\rm Earth}$ and initial lunar semi-major axis, $a_0$. $k_{2,{\rm Earth}}/Q_{\rm Earth}$ was held constant through orbital evolution in each case. For tidal properties typical of solid bodies ($k_{2,{\rm Earth}}/Q_{\rm Earth}>10^{-3}$), the transition occurs within the first million years of lunar orbital evolution.
}
\label{sup:fig:k2Q_dependence_transition}
\end{figure}

The transition between the early and late time regimes occurs around a transitional time of
\begin{equation}
\label{sup:eqn:tlimit_rev}
    t_{\rm tran} = \frac{2}{39} \frac{Q_{\rm Earth}}{k_{2,{\rm Earth}}} \frac{M_{\rm Earth}}{M_{\rm Moon}\sqrt{G(M_{\rm Earth}+M_{\rm Moon})}} \left( \frac{1}{R_{\rm Earth}} \right )^5  a_0^{13/2} \, ,
\end{equation}
which is a function of $k_{2,{\rm Earth}}/Q_{\rm Earth}$ and $a_0$ only. The crosses and dots in Figure~\ref{sup:fig:k2Q_dependence}b-d show the $k_{2,{\rm Earth}}/Q_{\rm Earth}$, and corresponding deformation rate, at which $t_{\rm tran}$ and $10$~$t_{\rm tran}$ are exceeded for each $a_0$ and $t$. By $10$~$t_{\rm tran}$ the $k_{2,{\rm Earth}}/Q_{\rm Earth}$ dependence of the deformation rate is firmly in the late $t$, weakly $k_{2,{\rm Earth}}/Q_{\rm Earth}$ dependent, regime. Figure~\ref{sup:fig:k2Q_dependence_transition} shows the transitional time as a function of $k_{2,{\rm Earth}}/Q_{\rm Earth}$ and $a_0$. For all but the lowest $k_{2,{\rm Earth}}/Q_{\rm Earth}$ cases, by 1~Myr after Moon formation the deformation rate is in the late $t$ limit. For bodies with tidal dissipation rates typical of solid bodies ($k_{2,{\rm Earth}}/Q_{\rm Earth}\sim$1$\times 10^{-2}$), the late $t$ limit is reached within 100~kyr. As the first crust on Earth is expected to form, at the very least, several thousand years after the Moon-forming impact \cite{Elkins-Tanton2008,Lebrun2013,Miyazaki2019a,Korenaga2023_rapid_solidification}, crustal deformation rates are likely only weakly dependent on tidal parameters for most of the period relevant for deformation of Earth's crust.

It is important to note that the above relations only hold if $k_{2,{\rm Earth}}/Q_{\rm Earth}$ does not vary significantly with time. However, as discussed in Section~\ref{sup:sec:tidal_model}, Earth's tidal properties were likely changing by many orders of magnitude during the early period of Earth's evolution. I have used the same simple tidal model described by Equation~\ref{sup:eqn:adot} to determine the effect of the increasing $k_{2,{\rm Earth}}/Q_{\rm Earth}$ of early Earth from that of a fluid planet ($\sim$10$^{-5}$) to that of the present-day Earth ($\sim$0.03). Figure~\ref{sup:fig:increasing_k2Q} shows the equatorial longitudinal convergence rates for cases where $k_{2,{\rm Earth}}/Q_{\rm Earth}$ increased either linearly (dashed lines) or exponentially (solid lines) with time such that the maximum $k_{2,{\rm Earth}}/Q_{\rm Earth}$ is reached at time, $t_{\rm max}$ (line colors). The time evolution of $k_{2,{\rm Earth}}/Q_{\rm Earth}$ in each case is shown in Figure~\ref{sup:fig:k2Q_functions}. At early times, the deformation rate at a given time is intermediate between cases where $k_{2,{\rm Earth}}/Q_{\rm Earth}$ was held constant with the minimum and maximum $k_{2,{\rm Earth}}/Q_{\rm Earth}$ (black solid lines in Figure~\ref{sup:fig:increasing_k2Q}). Within about 0.1~$t_{\rm max}$, the deformation rate exceeds that of the constant $k_{2,{\rm Earth}}/Q_{\rm Earth}$ cases, with a peak deformation rate at $t_{\rm max}$ where the deformation rate can be several times higher than the constant $k_{2,{\rm Earth}}/Q_{\rm Earth}$ cases. The deformation rate then tends to the deformation rate of the high constant $k_{2,{\rm Earth}}/Q_{\rm Earth}$ case from above. The crustal deformation rate at later times could therefore have been several times higher than estimated using the constant $k_{2,{\rm Earth}}/Q_{\rm Earth}$ model. However, given the orders of magnitude increase in $k_{2,{\rm Earth}}/Q_{\rm Earth}$, the difference in deformation rates between models with constant and varying tidal properties is still surprisingly modest. The largest increase in $k_{2,{\rm Earth}}/Q_{\rm Earth}$ is thought to have occurred when a substantial fraction of the mantle reached the rheological transition \cite{Zahnle2015}, and it is expected that there was a substantial enhancement in deformation around the time of the formation of the first crust. The coincidence of increased deformation rates and crustal formation could have interesting implications for the formation process and structure of the primary crust. 

\begin{figure}
\centering
\includegraphics[width=0.9\textwidth]{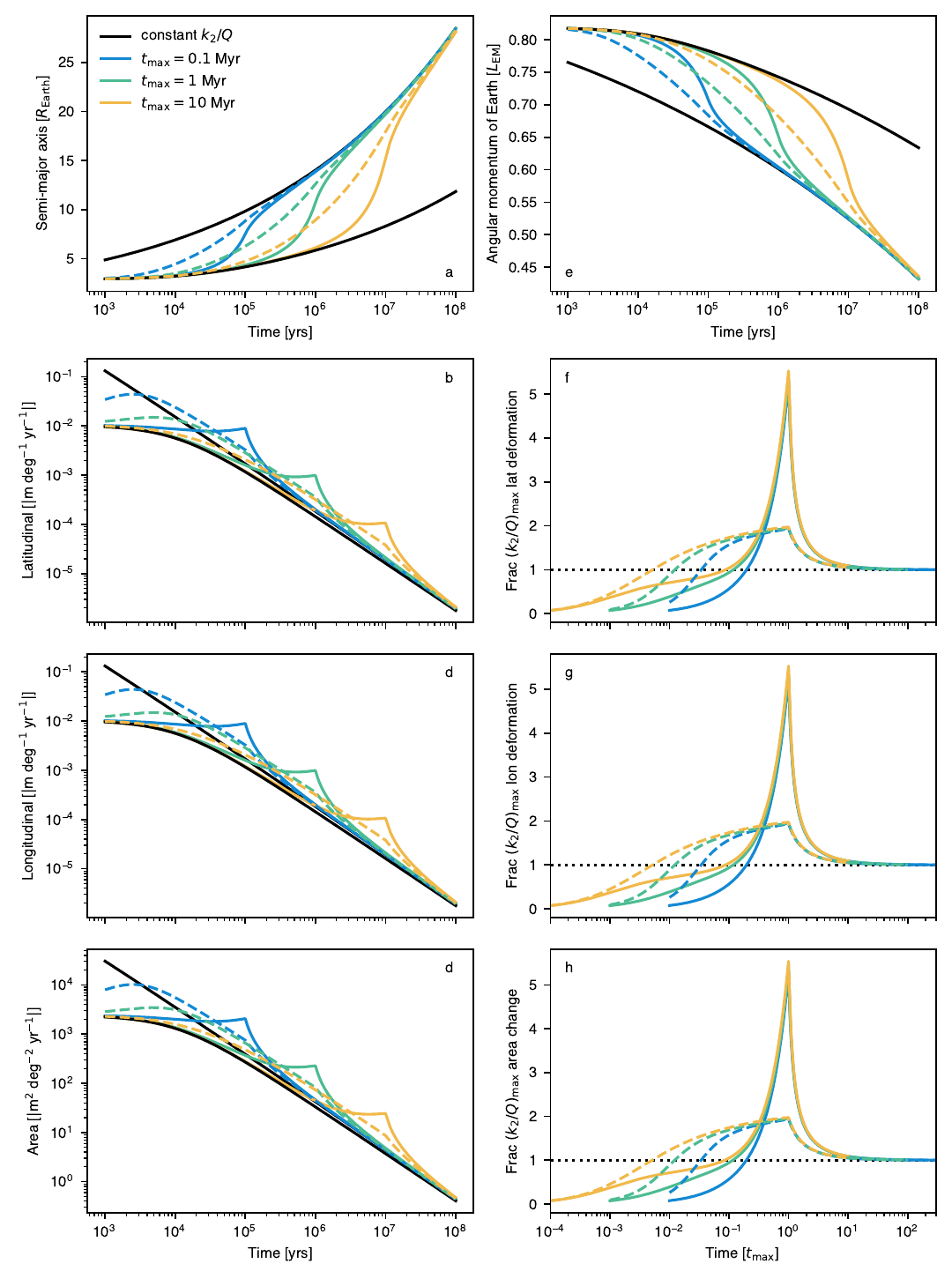}
\caption{Caption on next page.}
\end{figure}
\addtocounter{figure}{-1}
\begin{figure}
\centering
\caption{\add{Orbital evolutions and crustal deformation rates for tidal evolution models with varying tidal parameters. }The expected increase in $k_{2,{\rm Earth}}/Q_{\rm Earth}$ early in Earth's history \add{associated with magma ocean crystalization (see Section}~\ref{sup:sec:tidal_model:description}\add{) }have led to an increase in crustal deformation rates at later times compared to constant $k_{2,{\rm Earth}}/Q_{\rm Earth}$ models. Shown are the semi-major axis (a), rate of change of local length in the latitudinal (b) and longitudinal (c) directions, the rate of change in local surface area (d), and Earth's angular momentum (e) at varying times (colors) for tidal models with different $k_{2,{\rm Earth}}/Q_{\rm Earth}$ evolutions. Deformation rates are shown at the equator. Colored lines show the evolution when $k_{2,{\rm Earth}}/Q_{\rm Earth}$ increases exponentially (solid lines) or linearly (dashed lines) over different times periods (varying $t_{\rm max}$, colors). The time evolution of $k_{2,{\rm Earth}}/Q_{\rm Earth}$ in each case is shown in Figure~\ref{sup:fig:k2Q_functions}. Black lines show the result for scenarios where $k_{2,{\rm Earth}}/Q_{\rm Earth}$ was held constant at the minimum and maximum $k_{2,{\rm Earth}}/Q_{\rm Earth}$ used in the time dependent $k_{2,{\rm Earth}}/Q_{\rm Earth}$ examples\add{, 10$^{-5}$ and $3\times10^{-3}$ (that of the present-day solid Earth), respectively.} Panels f--h show the relative magnitude of the deformation rates in the time-dependent $k_{2,{\rm Earth}}/Q_{\rm Earth}$ and constant $k_{2,{\rm Earth}}/Q_{\rm Earth}=\left ( k_{2,{\rm Earth}}/Q_{\rm Earth} \right )_{\rm max}$\add{ and $k_{2,{\rm Earth}}/Q_{\rm Earth}=\left ( k_{2,{\rm Earth}}/Q_{\rm Earth} \right )_{\rm min}$ cases}. In panels f--h the time is normalized to the time at which the maximum $k_{2,{\rm Earth}}/Q_{\rm Earth}$ was reached in each scenario ($t_{\rm max}$). The black dotted line indicates when the crustal deformation rates in the time-dependent and constant $k_{2,{\rm Earth}}/Q_{\rm Earth}$ models are equal.
}
\label{sup:fig:increasing_k2Q}
\end{figure}

\section{\add{Comparison of tidal parameters used in Rufu \& Canup (2020) and other models}}
\label{sup:sec:tidal_params}

\citeA{Rufu2020} used different parameters determine to control the rate of orbital evolution than in the two other models considered in this work. Here, I give the relationship between the different sets of parameters.

\citeA{Rufu2020} used the tidal $A$ parameter, defined as
\begin{equation}
   A= \frac{k_{2, {\rm Moon}}}{k_{2, {\rm Earth}}} \frac{\Delta t_{\rm Moon}}{\Delta t_{\rm Earth}}\left(\frac{M_{\rm Earth}}{M_{\rm Moon}}\right)^{2}\left(\frac{R_{\rm Moon}}{R_{\rm Earth}}\right)^{5} \; ,
\end{equation}
 and the tidal time constant, defined as
\begin{equation}
t_T=\frac{1}{6 k_{2,{\rm Earth}} \mu \Omega^2_{\rm Earth} \Delta t_{\rm Earth}} \;.
 \end{equation}
$k_{2, {\rm Earth}}$ and $k_{2, {\rm Moon}}$ are the tidal love numbers of the Earth and Moon, respectively; $\Delta t_{\rm Earth}$ and $\Delta t_{\rm Moon}$ are the terrestrial and lunar tidal time delays, respectively; $M_{\rm Earth}$ and $M_{\rm Moon}$ are the mass of the Earth and Moon; $R_{\rm Earth}$ and $R_{\rm Moon}$ are the radius of the Earth and Moon; $\mu=M_{\rm Moon}/M_{\rm Earth}$; and
\begin{equation}
    \Omega_{\rm Earth}=\sqrt{\frac{G M_{\rm Earth}}{R^3_{\rm Earth}}} \; ,
\end{equation}
where $G$ is the gravitational constant. $t_T$ can be related to an effective $k_{2,{\rm Earth}}/Q_{\rm Earth}$ as
\begin{equation}
    \frac{k_{2,{\rm Earth}}}{Q_{\rm Earth}}=\frac{R^3_{\rm Earth}}{3GM_{\rm Moon}t_T} \left ( s_{0, {\rm Earth}} -\sqrt{\frac{G M_{\rm Earth}}{a^3_0}} \right ) \; ,
\end{equation}
where $a_0=3.5 R_{\rm Earth}$ and $s_{0, {\rm Earth}}$ are the initial semi-major axis and spin rate of Earth in the simulation. 

%



%
%

\section*{Open Research}
All planetary structure results presented in this study were produced using the HERCULES code v1 available through Zenodo \cite{Lock2019HERCULES} and the GitHub repository: \newline\noindent{\small {\tt https://github.com/sjl499/HERCULESv1\_user}}. The equations of state used, and relevant documentation, are available though Zenodo \cite{Stewart2020ironEOS,Stewart2019forsteriteEOS}. Tidal evolution results were either calculated using the analytical models described in Methods or were provided by authors of previous works \cite{Cuk2021,Rufu2020}. The scripts used to perform the calculations and produce the figures for this work will be made publicly available through GitHub and Zenodo upon acceptance.

\acknowledgments
I would like to thank Sarah Stewart, Paul Asimow, Mike Gurnis, Harriet Lau, Mathew Roche, and Matija \'{C}uk for useful discussions, and Raluca Rufu for providing the results of their simulations. This work was supported in part by the Division of Geological and Planetary Sciences at the California Institute of Technology, the US NSF (awards EAR-1947614 and EAR-1725349), and the UK NERC (grant NE/V014129/1). \\

\noindent \textbf{Author Contributions:} SJL conceived the project, performed simulations, interpreted the results, and wrote the paper. \\

\noindent \textbf{Competing Interests:} The author declares that they have no competing financial interests.

%
\bibliography{References_processed}
%


%
%
%
%
%

\end{document}